\documentclass[12pt]{article}
\usepackage{natbib}
\usepackage{graphicx}
\usepackage{amssymb,amsfonts,amsmath}
\usepackage{geometry}
\usepackage{setspace}

\usepackage{multirow}

\begin{document}
\title{Flexible Tweedie regression models for continuous data}
\author{Wagner H. Bonat\thanks{Department of Mathematics and Computer Science, University of Southern Denmark, Odense, Denmark.
		    Department of Statistics, Paran\'a Federal University, 
		    Curitiba, Paran\'a, Brazil. E-mail: wbonat@ufpr.br} 
~and C\'elestin C. Kokonendji\thanks{Universit\'e de Franche-Comt\'e, Laboratoire de Math\'ematiques de Besan\c{c}on, 
		    Besan\c{c}on, France. E-mail:celestin.kokonendji@univ-fcomte.fr}}
\date{\vspace{-5ex}}
\maketitle

\begin{abstract}
Tweedie regression models provide a flexible family of distributions to deal 
with non-negative highly right-skewed data as well as symmetric and 
heavy tailed data and can handle continuous data with probability mass at zero.
The estimation and inference of Tweedie regression 
models based on the maximum likelihood method are challenged by the presence 
of an infinity sum in the probability function and non-trivial restrictions 
on the power parameter space.
In this paper, we propose two approaches for fitting Tweedie regression 
models, namely, quasi- and pseudo-likelihood.
We discuss the asymptotic properties of the two approaches and perform 
simulation studies to compare our methods with the maximum 
likelihood method. In particular, we show that the quasi-likelihood method
provides asymptotically efficient estimation for regression parameters.
The computational implementation of the alternative methods is faster and easier than
the orthodox maximum likelihood, relying on a simple Newton scoring algorithm.
Simulation studies showed that the quasi- and pseudo-likelihood approaches present estimates,
standard errors and coverage rates similar to the maximum likelihood method.
Furthermore, the second-moment assumptions required by the quasi- and pseudo-likelihood
methods enables us to extend the Tweedie regression models to the class of quasi-Tweedie
regression models in the Wedderburn's style. Moreover, it allows to eliminate the non-trivial 
restriction on the power parameter space, and thus provides a flexible 
regression model to deal with continuous data.
We provide \texttt{R} implementation and illustrate the 
application of Tweedie regression models using three data sets.
\end{abstract}

\section{Introduction} \label{Intro}

Statistical modelling is one of the most important areas of applied 
statistics with applications in many fields of scientific research, 
such as sociology, economy, ecology, agronomy, insurance, medicine 
to cite but a few. 
There exists an infinity of statistical modelling frameworks, 
but the class of Generalized Linear Models (GLM)~\citep{Nelder:1972} 
is the most used in the last four decades. 
The success of this approach is due to its ability to deal with different 
types of response variables, such as binary, count and continuous in a 
general framework with a powerful scheme for estimation and inference 
based on the likelihood paradigm.

Special cases of the GLM class include the Gaussian linear model to deal 
with continuous data, gamma and inverse Gaussian regression models for 
handling positive continuous data. Logistic and Poisson regression models for 
dealing with binary or binomial and count data, respectively. 
These models are linked, since they belong to the 
class of the exponential dispersion models~\citep{Jorgensen:1987,Jorgensen:1997}, 
and share the property to be described by their first two moments, 
mean and variance. Furthermore, the variance function plays an important 
role in the context of exponential dispersion models, since it describes 
the relationship between the mean and variance and characterizes the 
distribution~\citep{Jorgensen:1997}.

Let $Y$ denote the response variable and assume that the density probability 
function of $Y$ belongs to the class of exponential dispersion models. 
Furthermore, we assume that $\mathrm{E}(Y) = \mu$ and 
$\mathrm{Var}(Y) = \phi V(\mu) = \phi \mu^p$ then 
$Y \sim \mathrm{Tw_p}(\mu, \phi)$, where $\mathrm{Tw_p}(\mu, \phi)$ denotes a 
Tweedie~\citep{Tweedie:1984,Jorgensen:1997} random variable with mean 
$\mu$ and variance $\phi \mu^p$, such that $\phi > 0$ and 
$p \in (-\infty ,0] \cup [1,\infty)$ are the dispersion and power 
parameters, respectively. 
The support of the distribution depends on the value of the power parameter. 
For $p \geq 2$, $1 < p < 2$ and $p = 0$ the support corresponds to the positive, 
non-negative and real values, respectively. 
In these cases $\mu \in \Omega$, where $\Omega$ is the 
convex support (i.e. the interior of the closed convex hull of the 
corresponding distribution support). Finally, for $p < 0$ the support 
corresponds to the real values, however the expectation $\mu$ is positive.

For practical data analysis, the Tweedie distribution is interesting,
since it has as special cases the Gaussian ($p = 0$), Poisson ($p = 1$), 
non-central gamma ($p=3/2$), gamma ($p = 2$) and inverse Gaussian ($p = 3$) 
distributions~\citep{Jorgensen:1987,Jorgensen:1997}.
Another important case often applied in the context of insurance data
\citep{Smyth:2002,Jorgensen:1994} corresponds to the 
compound Poisson distribution, obtained when $1 < p < 2$. 
The compound Poisson distribution is a frequent choice for the modelling 
of non-negative data with probability mass at zero and highly right-skewed.

The power parameter plays an important role in the context of
Tweedie models, since it is an index which distinguishes between some 
important continuous distributions. The algorithms we shall propose
in Section~\ref{estimation} allow us to estimate the power parameter, 
which works as an automatic distribution selection.
Although, the estimation of the regression parameters is less affected by 
the dispersion structure, the standard errors associated with the regression 
parameters are determined by dispersion structure, which justifies dedicate 
attention to the estimation of the power and dispersion parameters.

The orthodox approach is based on the likelihood paradigm, which in turn 
is an efficient estimation method. 
However, a particularity about the Tweedie distribution is that outside 
the special cases, its probability density function cannot be written in 
a closed form, and requires numerical methods for evaluating the density 
function. \citet{Dunn:2005,Dunn:2008} proposed methods to evaluate the 
density function of the Tweedie distribution, but these methods are 
computationally demanding and show different levels of accuracy for 
different regions of the parameter space. 
Furthermore, the parameter space associated with the power parameter 
presents non-trivial restrictions. Current software 
implementations~\citep{tweedie:2013} are restricted to dealing with $p \geq 1$.
These facts become the process of inference based on the likelihood paradigm 
difficult and sometimes slow.

The main goal of this paper is to propose alternative methods for 
estimation and inference of Tweedie regression models. 
In particular, we discuss the quasi-likelihood \citep{Jorgensen:2004, Bonat:2016} 
and pseudo-likelihood~\citep{Gourieroux:1984} approaches.
These methods are fast and simple computationally, because they employ 
the first two moments, merely avoiding to evaluate the probability density 
function. Moreover, the second-moment assumptions required by the 
quasi- and pseudo-likelihood methods allow us to extend the Tweedie 
regression models to the class of quasi-Tweedie regression models 
in the style of~\citet{Wedderburn:1974}. 
The weaker assumptions of the second-moments specification eliminate the 
restrictions on the parameter space of the power parameter.
Hence, it is possible to estimate negative and between zero and one values 
for the power parameter. In this way, we overcome the main restrictions 
of current software implementations and provide a flexible 
regression model to deal with continuous data.

We present the theoretical development of the quasi- and pseudo-likelihood 
methods in the context of Tweedie regression models. 
In particular, we show analytically that the quasi-likelihood 
approach provides asymptotic efficient estimation for regression parameters.
We present efficient and stable fitting algorithms based on the
two new approaches and provide \texttt{R} computational implementation. 
We employed simulation studies to compare the properties of
our approaches with the maximum likelihood method in a finite sample scenario.
We compare the approaches in terms of bias, efficiency and coverage rate of the
confidence intervals. Furthermore, we explore the flexiblity of Tweedie
regression models to deal with heavy tailed distributions.

Tweedie distributions are extensively used in statistical modelling, 
thereby motivating the study of their estimation in a more general framework.
Applications include \citet{Lee:1993,Neilsen:2001,Vinogradov:2004},
who applied Tweedie distributions for describing the chaotic
behaviour of stock price movements. Further applications include property and causality 
insurance, where \citet{Jorgensen:1994} and \citet{Smyth:2002} fit the Tweedie 
family to automobile insurance claims data. 
Tweedie distributions have also found applications in 
biology~\citep{Kendal:2004, Kendal:2007}, 
fisheries research \citep{Foster:2013,Shono:2008},
genetics and medicine~\citep{Kendal:2000}. 
\citet{Chen:2010} presented Bayesian semiparametric models based on the
reproductive form of exponential dispersion models.
\citet{Zhang:2013} discussed the maximum likelihood and Bayesian estimation
for Tweedie compound Poisson linear mixed models. 
For a recent application and further references see \citet{Bonat:2016}. 

The rest of the paper is organized as follows. 
In the next section, we provide some background about Tweedie regression models. 
Section $3$ discusses the 
approaches to estimation and inference. 
Section $4$ presents the main results from our simulation studies. 
Section $5$ presents the application of Tweedie regression models 
to three data sets. The first one concerns daily precipitation in Curitiba, 
Paran\'a State, Brazil. This dataset illustrates the analysis of positive
continuous data with probability mass at zero. The second data set corresponds
to a cross-section study developed for studying the income dynamics in Australia.
This dataset shows the analysis of positive, highly right-skewed response variable.
The last data set illustrates the analysis of symmetric positive data, where
current implementations have problems to deal with power parameter smaller than $1$.
Finally, Section $6$ reports some final remarks. The \texttt{R} implementation
is available in the supplementary material.

\section{Tweedie regression models}

The Tweedie distribution belongs to the class of exponential dispersion 
models (EDM) \citep{Jorgensen:1987,Jorgensen:1997}. Thus, for a random variable $Y$ which 
follows an EDM, the density function can be written as:
\begin{equation*}
\label{distri}
f_{Y}(y; \mu, \phi, p) = a(y,\phi,p) \exp\{(y\psi - k(\psi))/\phi\},
\end{equation*}
where $\mu = \mathrm{E}(Y) = k^{\prime}(\psi)$ is the mean, 
$\phi > 0$ is the dispersion parameter, $\psi$ is the canonical 
parameter and $k(\psi)$ is the cumulant function. 
The function $a(y,\phi, p)$ cannot be written in a closed form apart of the 
special cases cited. The variance is given by $\mathrm{Var}(Y) = \phi V(\mu)$ 
where $V(\mu) = k^{\prime \prime}(\psi)$ is called the variance function. 
Tweedie densities are characterized by power variance functions of the 
form $V(\mu) = \mu^p$, where $p \in (-\infty  ,0] \cup [1,\infty)$ is 
the index determining the distribution. 
Although, Tweedie densities are not known in closed form, their cumulant 
generating function is simple. The cumulant generating function is given by
\begin{equation*}
K(t) = \{ k(\psi + \phi t) - k(\psi) \} / \phi,
\end{equation*}
where $k(\psi)$ is the cumulant function,
\begin{equation*}
\psi = \left\{ \begin{matrix}
\frac{\mu^{1-p}}{1-p} & p \neq 1 \\ 
 \log \mu & p = 1 
\end{matrix} \right.
\quad
\text{and}
\quad
k(\psi) = \left\{\begin{matrix}
\frac{\mu^{2-p}}{2-p} & p \neq 2 \\ 
 \log \mu & p = 2. 
\end{matrix}  \right.
\end{equation*}

The remaining factor in the density, $a(y,\phi, p)$ needs to be evaluated 
numerically. \citet{Jorgensen:1997} presents two series expressions for 
evaluating the density, for $1 < p < 2$ and for $p > 2$. 
In the first case can be shown that,
\begin{equation*}
P(Y = 0) = \exp \left \{ -\frac{\mu^{2-p}}{\phi(2-p)} \right \}
\end{equation*}
and for $y > 0$ that
\begin{equation*}
a(y,\phi,p) = \frac{1}{y} W(y,\phi,p),
\end{equation*}
with $W(y,\phi,p) = \sum_{k=1}^{\infty} W_k$ and
\begin{equation*}
W_k = \frac{y^{-k\alpha}(p-1)^{\alpha k}}{\phi^{k(1-\alpha)}(2 - p)^k k! \Gamma(-k\alpha)},
\end{equation*}
where $\alpha = (2 -p)/(1-p)$.

A similar series expansion exists for $p > 2$ and it is given by:
\begin{equation*}
a(y,\phi,p) = \frac{1}{\pi y} V(y, \phi, p),
\end{equation*}
with $V = \sum_{k=1}^{\infty} V_k$ and
\begin{equation*}
V_k = \frac{\Gamma(1 + \alpha k) \phi^{k(\alpha - 1)} (p - 1)^{\alpha k}}{\Gamma(1+k) (p -2)^k y^{\alpha k}} (-1)^k \sin(-k \pi \alpha).
\end{equation*}

\citet{Dunn:2005} presented detailed studies about these series and an 
algorithm to evaluate the Tweedie density function based on series expansions. 
The algorithm is implemented in the package \texttt{tweedie}~\citep{tweedie:2013} 
for the statistical software \texttt{R}\citep{R:2016} through the function 
\texttt{dtweedie.series}. \citet{Dunn:2008} also studied 
two alternative methods to evaluate the density function of the Tweedie 
distributions, one based on the inversion of cumulant generating 
function using the Fourier inversion and the sandlepoint approximation, 
for more details see \citet{tweedie:2013}. 
In this paper, we used only the approach described in this Section, i.e.
based on series expansions.

We now turn to Tweedie regression models. 
Consider a cross-sectional dataset, $(y_i, \boldsymbol{x}_i)$, $i = 1, \ldots, n$, where
$y_i$'s are i.i.d. realizations of $Y_i$ according to
$Y_i \sim \mathrm{Tw_p}(\mu_i, \phi)$ and
$g(\mu_i) = \eta_i = \boldsymbol{x}_i^{\top} \boldsymbol{\beta}$,
where $\boldsymbol{x}_i$ and $\boldsymbol{\beta}$ are ($Q \times 1$) vectors of 
known covariates and unknown regression parameters, respectively. 
It is straightforward to see that 
$\mathrm{E}(Y_i) = \mu_i = g^{-1}(\boldsymbol{x}_i^{\top} \boldsymbol{\beta})$ and the 
$\mathrm{Var}(Y_i) = C_i = \phi \mu_i^p$. 
Hence, the model is equivalently specified by its joint distribution 
and by its first two moments. The Tweedie regression model is parametrized by 
$\boldsymbol{\theta} = (\boldsymbol{\beta}^\top, \boldsymbol{\lambda}^\top = (\phi = \exp(\delta), p)^\top)^\top$.
Note that, we introduce the reparametrization $\phi = \exp(\delta)$ for computational convenience. 
Finally, in this paper we adopt the orthodox logarithm link function.

\section{Estimation and Inference}\label{estimation}
This section is devoted to estimation and inference of Tweedie regression models.
In what follows, we shall discuss the maximum likelihood, quasi-likelihood and
pseudo-likelihood methods.

\subsection{Maximum likelihood estimation}

The maximum likelihood estimator (MLE) for the parameter vector 
$\boldsymbol{\theta}$ denoted by $\boldsymbol{\hat{\theta}_M}$ 
is obtained by maximizing the following log-likelihood function,
\begin{equation}
\label{loglikelihood}
\mathcal{L}(\boldsymbol{\theta})=\sum^n_{i=1}\log\left \{ a(y_i;\boldsymbol{\lambda}) \right \}+ \frac{1}{\exp(\delta)}(y_i\psi_i-k(\psi_i)).
\end{equation}
As we shall show below the vectors $\boldsymbol{\beta}$ and $\boldsymbol{\lambda}$ are orthogonal,
hence is sensible to discuss each of them separately.
The score function for the regression parameters $\boldsymbol{\beta} = (\beta_0, \ldots, \beta_Q)$ is given by
\begin{equation*}
\mathcal{U}_{\boldsymbol{\beta}}(\boldsymbol{\beta}, \boldsymbol{\lambda}) = \left ( \frac{\partial \mathcal{L}(\boldsymbol{\theta})}{\partial \beta_1}^\top, \ldots, \frac{\partial \mathcal{L}(\boldsymbol{\theta})}{\partial \beta_Q}^\top \right )^\top,
\end{equation*}
where
\begin{eqnarray}
\frac{\partial \mathcal{L}(\boldsymbol{\theta})}{\partial \beta_j} &=& \sum_{i=1}^n \frac{\partial \mathcal{L}(\boldsymbol{\theta})}{\partial\psi_i}\frac{\partial\psi_i}{\partial\mu_i}\frac{\partial\mu_i}{\partial\eta_i}\frac{\partial\eta_i}{\partial\beta_j} \nonumber \\
&=& \sum^n_{i=1}\mu_i x_{ij}\left [ \frac{1}{\exp(\delta)\mu_i^p} \right ](y_i-\mu_i), \quad \text{for} \quad j = 1, \ldots,Q. \nonumber
\end{eqnarray}
The entry $(j,k)$ of the $Q \times Q$ Fisher information matrix $\mathcal{F}_{\boldsymbol{\beta}}$ for the regression coefficients is given by
\begin{equation}
\label{fisherbeta}
\mathcal{F}_{\boldsymbol{\beta}_{jk}} =-\mathrm{E} \left \{ \frac{\partial^2 \mathcal{L}(\boldsymbol{\theta})}{\partial\beta_j\partial\beta_k} \right \}=\sum^n_{i=1}\mu_i x_{ij}\left [ \frac{1}{\exp(\delta)\mu_i^p} \right ]\mu_i x_{ik}.
\end{equation}

Similarly, the score function for the dispersion parameters $\boldsymbol{\lambda} = (\exp(\delta), p)$ is given by
\begin{equation*}
\mathcal{U}_{\boldsymbol{\lambda}}(\boldsymbol{\lambda}, \boldsymbol{\beta}) = \left ( \frac{\partial \mathcal{L}(\boldsymbol{\theta})}{\partial \delta}^\top, \frac{\partial \mathcal{L}(\boldsymbol{\theta})}{\partial p}^\top \right )^\top,
\end{equation*}
whose components are given by
\begin{equation}
\label{derphi}
\frac{\partial \mathcal{L}(\boldsymbol{\theta})}{\partial \delta} = \sum^n_{i=1}\frac{\partial}{\partial\delta}\log a(y_i;\boldsymbol{\lambda})-\frac{1}{\exp(\delta)}(y_i\psi_i-\kappa(\psi_i))
\end{equation}
and
\begin{equation}
\label{derp}
\frac{\partial \mathcal{L}(\boldsymbol{\theta})}{\partial p} = \sum^n_{i=1}\frac{\partial}{\partial p}\log a(y_i;\boldsymbol{\lambda})+\frac{1}{\exp(\delta)}\left [ y_i\frac{\partial\psi_i}{\partial p} - \frac{\partial\kappa(\psi_i)}{\partial p} \right  ].
\end{equation}

The entry $(j,k)$ of the $2 \times 2$ Fisher information matrix $\mathcal{F}_{\boldsymbol{\lambda}}$ for the dispersion parameters is given by
\begin{equation}
\label{fisherdisp}
\mathcal{F}_{\boldsymbol{\lambda}_{jk}} =-\mathrm{E} \left \{ \frac{\partial^2 \mathcal{L}(\boldsymbol{\theta})}{\partial\lambda_j\partial\lambda_k} \right \}.
\end{equation}

The derivative in equations (\ref{derphi}), (\ref{derp}) and (\ref{fisherdisp}) depends on the
derivative of the infinite sum $a(y_i;\boldsymbol{\lambda})$, and it cannot
be expressed in closed form. Hence, numerical methods are required for approximating these derivatives. 
Let $\tilde{\mathcal{U}}_{\boldsymbol{\lambda}}$ and $\tilde{\mathcal{F}}_{\boldsymbol{\lambda}}$ denote
the approximated score function and observed information matrix for the dispersion parameters, respectively.
In this paper, we adopted the Richardson method \citep{Richard:1994}, as implemented in the \texttt{R} package 
\texttt{numDeriv}~\citep{rootsolve:2015} for computing these approximations.
Furthermore, the cross entries of the Fisher information matrix are given by
\begin{equation*}
\mathcal{F}_{\beta_j\delta}= -\mathrm{E} \left \{ \frac{\partial \mathcal{U}_{\beta_j}(\boldsymbol{\beta},\boldsymbol{\lambda})}{\partial \delta} \right \}=-\mathrm{E}\left \{ \mu_i x_{ij}\left [-\frac{1}{\exp(\delta)\mu_i^p} \right ](y_i-\mu_i) \right \} = 0 
\end{equation*}
and
\begin{equation*}
\mathcal{F}_{\beta_j p}= -\mathrm{E} \left \{ \frac{\partial \mathcal{U}_{\beta_j}(\boldsymbol{\beta},\boldsymbol{\lambda})}{\partial p} \right \}= -\mathrm{E} \left \{ \mu_i x_{ij} \left [ \frac{\partial}{\partial p}\frac{1}{\exp(\delta)\mu_i^p} \right ](y_i-\mu_i) \right \} = 0.
\end{equation*}
Hence, the vectors $\boldsymbol{\beta}$ and $\boldsymbol{\lambda}$ are orthogonal. The joint Fisher information matrix for $\boldsymbol{\theta}$ is given by
\begin{equation*}
\mathcal{F}_{\boldsymbol{\theta}} = \begin{pmatrix}
\mathcal{F}_{\boldsymbol{\beta}} & \boldsymbol{0} \\ 
\boldsymbol{0} & \mathcal{F}_{\boldsymbol{\lambda}}
\end{pmatrix},
\end{equation*}
whose entries are defined by (\ref{fisherbeta}) and (\ref{fisherdisp}).
Finally, the asymptotic distribution of $\boldsymbol{\hat{\theta}_M}$ is
\begin{equation*}
\boldsymbol{\hat{\theta}_M} \sim \mathrm{N}(\boldsymbol{\theta}, \mathcal{F}_{\boldsymbol{\theta}}^{-1})
\end{equation*} 
where $\mathcal{F}_{\boldsymbol{\theta}}^{-1}$ denote the inverse of the Fisher information matrix.
In practice the entry $\mathcal{F}_{\boldsymbol{\lambda}}$ is replaced by the approximation $\tilde{\mathcal{F}}_{\boldsymbol{\lambda}}$.

In order to solve the system of equations $\mathcal{U}_{\boldsymbol{\beta}} = \boldsymbol{0}$ and $\tilde{\mathcal{U}}_{\boldsymbol{\lambda}} = \boldsymbol{0}$, we employ the two steps Newton scoring algorithm, defined by
\begin{eqnarray}
\label{NewtonScoring}
\boldsymbol{\beta}^{(i+1)} &=& \boldsymbol{\beta}^{(i)} - \mathcal{F}_{\boldsymbol{\beta}}^{-1} \mathcal{U}_{\boldsymbol{\beta}}(\boldsymbol{\beta}^{(i)}, \boldsymbol{\lambda}^{(i)}) \nonumber \\
\boldsymbol{\lambda}^{(i+1)} &=& \boldsymbol{\lambda}^{(i)} - \tilde{\mathcal{F}}_{\boldsymbol{\lambda}}^{-1} \tilde{\mathcal{U}}_{\boldsymbol{\lambda}}(\boldsymbol{\beta}^{(i+1)}, \boldsymbol{\lambda}^{(i)}),
\end{eqnarray}
which in turn explicitly uses the orthogonality between $\boldsymbol{\beta}$ and $\boldsymbol{\lambda}$.

The numerical evaluation of the derivatives required in equations (\ref{derphi}), (\ref{derp}) and (\ref{fisherdisp})
can be inaccurate, mainly for $p \approx 1$, i.e. the border of the parameter space. 
Thus, an alternative approach is to maximize directly the log-likelihood
function in equation (\ref{loglikelihood}) using a derivative-free algorithm as the Nelder-Mead method~\citep{Nelder:1965}.
A more computationally efficient approach is to use the Nelder-Mead algorithm for maximizing only the 
profile log-likelihood for the dispersion parameters, which in turn is obtained by inserting
the first equation of the two steps Newton scoring algorithm (\ref{NewtonScoring}) 
in the log-likelihood function (\ref{loglikelihood}).
Note that, by using this approach for each evaluation of the profile likelihood, 
we have a maximization problem for the regression parameters. 
We implemented these three approaches to obtain the maximum likelihood estimator.
The direct maximization of the log-likelihood function using the Nelder-Mead algorithm is
slow, mainly for large number of regression coefficients. The two steps Newton
scoring algorithm presented many convergence problems for small values of the power
parameter. Finally, the profile likelihood approach is the fast and stable implementation.
However, the profile likelihood approach presented problems to compute the
standard errors associated with the dispersion estimates for $p \approx 1$.
In this paper, we used only the approach based on the profile log-likelihood, but we
also provide \texttt{R} code for the other two approaches.

\subsection{Quasi-likelihood estimation}

We shall now introduce the quasi-likelihood estimation using terminology and results
from \citet{Jorgensen:2004, Holst:2015, Bonat:2016}. The quasi-likelihood approach
adopted in this paper combines the quasi-score and Pearson estimating functions
to estimation of regression and dispersion parameters, respectively. The approach 
is also discussed in the context of estimating functions, see 
\citet{Liang:1995,Jorgensen:2004} for further details.

The quasi-score function for $\boldsymbol{\beta}$ has the following form,

\begin{equation*}
\mathcal{U}^{q}_{\boldsymbol{\beta}}(\boldsymbol{\beta}, \boldsymbol{\lambda}) = \left (\sum_{i=1}^n \frac{\partial \mu_i}{\partial \beta_1}C^{-1}_i(y_i - \mu_i)^\top, \ldots, \sum_{i=1}^n \frac{\partial \mu_i}{\partial \beta_Q}C^{-1}_i(y_i - \mu_i)^\top  \right )^\top,
\end{equation*}
where $\partial \mu_i/\partial \beta_j = \mu_i x_{ij}$ for $j = 1, \ldots, Q$.
The entry $(j,k)$ of the $Q \times Q$ sensitivity matrix for $\mathcal{U}^{q}_{\boldsymbol{\beta}}$ is given by

\begin{equation}
\label{Sbeta}
\mathrm{S}_{\beta_{jk}} = \mathrm{E}\left ( \frac{\partial}{\partial \beta_k} \mathcal{U}^{q}_{\beta_j}(\boldsymbol{\beta}, \boldsymbol{\lambda})  \right ) = -\sum_{i=1}^n \mu_i x_{ij} \left [ \frac{1}{\exp(\delta) \mu_i^p} \right ] x_{ik} \mu_i.
\end{equation}
In a similar way, the entry $(j,k)$ of the $Q \times Q$ variability matrix for $\mathcal{U}^{q}_{\boldsymbol{\beta}}$ is given by
\begin{equation*}
\label{Vbeta}
\mathrm{V}_{\beta_{jk}} = \mathrm{Var}(\mathcal{U}^{q}_{\boldsymbol{\beta}}(\boldsymbol{\beta}, \boldsymbol{\lambda})) = \sum_{i=1}^n \mu_i x_{ij} \left [ \frac{1}{\exp(\delta)\mu_i^p} \right ] x_{ik} \mu_i.
\end{equation*}

Following \citet{Jorgensen:2004, Bonat:2016}, the Pearson estimating function for the dispersion parameters has the following form,

\begin{equation*}
\label{Pearson}
\mathcal{U}^{q}_{\boldsymbol{\lambda}}(\boldsymbol{\lambda}, \boldsymbol{\beta}) = \left (\sum_{i=1}^n \boldsymbol{W}_{i\delta} \left [ (y_i - \mu_i)^2 - C_i \right ]^\top, \sum_{i=1}^n \boldsymbol{W}_{i p} \left [ (y_i - \mu_i)^2 - C_i \right ]^\top  \right )^\top,
\end{equation*}
where $\boldsymbol{W}_{i\delta} = - \partial C^{-1}_i/\partial \delta$ and $\boldsymbol{W}_{ip} = - \partial C^{-1}_i/\partial p$.
The Pearson estimating functions are unbiased estimating functions for $\boldsymbol{\lambda}$ 
based on the squared residuals $(y_i - \mu_i)^2$ with mean $C_i$. It is equivalent to treating the squared residual
as a gamma variable, which is hence close in spirit to Perry's gamma regression method \citep{Jorgensen:2011,Park:2004}.

We shall now calculate the sensitivity matrix for the dispersion parameters. The entry $(j,k)$ of the $2 \times 2$ sensitivity matrix is given by
\begin{equation*}
\mathrm{S}_{\boldsymbol{\lambda}_{jk}} = \mathrm{E}\left ( \frac{\partial}{\partial \lambda_k}\mathcal{U}^{q}_{\lambda_j}(\boldsymbol{\lambda}, \boldsymbol{\beta})  \right ) = -\sum_{i=1}^n \boldsymbol{W}_{i \lambda_j} C_i\boldsymbol{W}_{i\lambda_k}C_i, 
\end{equation*}
where $\lambda_1$ and $\lambda_2$ denote either $\delta$ or $p$, giving
\begin{equation}
\label{Slambda}
\mathrm{S}_{\boldsymbol{\lambda}} = \begin{pmatrix}
-n & -\sum_{i=1}^n\log(\mu_i) \\ 
-\sum_{i=1}^n \log(\mu_i) & -\sum_{i=1}^n \log(\mu_i)^2
\end{pmatrix}.
\end{equation}
Similarly, the cross entries of the sensitivity matrix are given by

\begin{equation}
\label{Sbetalambda}
\mathrm{S}_{\beta_j \lambda_k} = \mathrm{E}\left ( \frac{\partial}{\partial \lambda_k}\mathcal{U}^{q}_{\beta_j}(\boldsymbol{\beta}, \boldsymbol{\lambda})  \right ) = 0
\end{equation}
and
\begin{equation}
\label{Slambdabeta}
\mathrm{S}_{\lambda_j \beta_k} = \mathrm{E}\left ( \frac{\partial}{\partial \beta_k}\mathcal{U}^{q}_{\lambda_j}(\boldsymbol{\lambda}, \boldsymbol{\beta})  \right ) = -\sum_{i=1}^n \boldsymbol{W}_{i\lambda_j}C_i\boldsymbol{W}_{i\beta_k}C_i,
\end{equation}
where $\boldsymbol{W}_{i\beta_k} = -\partial C_i^{-1}/\partial \beta_k$. Finally, the joint sensitivity matrix for the parameter vector $\boldsymbol{\theta}$ is given by
\begin{equation*}
\mathrm{S}_{\boldsymbol{\theta}} = \begin{pmatrix}
\mathrm{S}_{\boldsymbol{\beta}} & \boldsymbol{0} \\ 
\mathrm{S}_{\boldsymbol{\lambda}\boldsymbol{\beta}} & \mathrm{S}_{\boldsymbol{\lambda}}
\end{pmatrix},
\end{equation*}
whose entries are defined by (\ref{Sbeta}), (\ref{Slambda}), (\ref{Sbetalambda}) and (\ref{Slambdabeta}).

We shall now calculate the asymptotic variance of the quasi-likelihood estimators denoted by $\boldsymbol{\hat{\theta}_{QL}}$, as
obtained from the inverse Godambe information matrix, whose general form is 
$J^{-1}_{\boldsymbol{\theta}} = \mathrm{S}^{-1}_{\boldsymbol{\theta}} \mathrm{V}_{\boldsymbol{\theta}} \mathrm{S}^{-\top}_{\boldsymbol{\theta}}$
for a vector of parameter $\boldsymbol{\theta}$, where $-\top$ denotes inverse transpose. 
The variability matrix for $\boldsymbol{\theta}$  has the form
\begin{equation}
\label{VTHETA}
\mathrm{V}_{\boldsymbol{\theta}} = \begin{pmatrix}
\mathrm{V}_{\boldsymbol{\beta}} & \mathrm{V}_{\boldsymbol{\beta}\boldsymbol{\lambda}} \\ 
\mathrm{V}_{\boldsymbol{\lambda}\boldsymbol{\beta}} & \mathrm{V}_{\boldsymbol{\lambda}}
\end{pmatrix},
\end{equation}
whereas $\mathrm{V}_{\boldsymbol{\lambda}\boldsymbol{\beta}} = \mathrm{V}^{\top}_{\boldsymbol{\beta}\boldsymbol{\lambda}}$ and $\mathrm{V}_{\boldsymbol{\lambda}}$ depend on the third and fourth moments of $Y_i$, respectively. In order to avoid this dependence on high-order moments, we propose to use the empirical versions of $\mathrm{V}_{\boldsymbol{\lambda}}$ and $\mathrm{V}_{\boldsymbol{\lambda}\boldsymbol{\beta}}$, which entries are given by

\begin{equation*}
\tilde{\mathrm{V}}_{\lambda_{jk}} = \sum_{i=1}^n \mathcal{U}^{q}_{\lambda_j}(\boldsymbol{\lambda}, \boldsymbol{\beta})_i\mathcal{U}^{q}_{\lambda_k}(\boldsymbol{\lambda}, \boldsymbol{\beta})_i \quad \text{and} \quad \tilde{\mathrm{V}}_{\lambda_j \beta_k} = \sum_{i=1}^n \mathcal{U}^{q}_{\lambda_j}(\boldsymbol{\lambda}, \boldsymbol{\beta})_i\mathcal{U}^{q}_{\beta_k}(\boldsymbol{\lambda}, \boldsymbol{\beta})_i.
\end{equation*}
Finally, the asymptotic distribution of $\boldsymbol{\hat{\theta}_{QL}}$ is given by
\begin{equation*}
\boldsymbol{\hat{\theta}_{QL}} \sim \mathrm{N}(\boldsymbol{\theta}, \mathrm{J}_{\boldsymbol{\theta}}^{-1}).
\end{equation*} 
We may show by using standard results for inverse of partitioned matrix that
\begin{equation*}
\mathrm{J}^{-1}_{\boldsymbol{\theta}} = \begin{pmatrix}
\mathrm{S}^{-1}_{\boldsymbol{\beta}} \mathrm{V}_{\boldsymbol{\beta}}\mathrm{S}^{-1}_{\boldsymbol{\beta}} & 
\mathrm{S}^{-1}_{\boldsymbol{\beta}}(-\mathrm{V}_{\boldsymbol{\lambda}}\mathrm{S}^{-1}_{\boldsymbol{\beta}}\mathrm{S}^{\top}_{\boldsymbol{\lambda}\boldsymbol{\beta}} + \mathrm{V}^{\top}_{\boldsymbol{\lambda}\boldsymbol{\beta}})\mathrm{S}^{-1}_{\boldsymbol{\lambda}} \\ 
\mathrm{S}^{-1}_{\boldsymbol{\lambda}}(- \mathrm{S}_{\boldsymbol{\lambda}\boldsymbol{\beta}}\mathrm{S}^{-1}_{\boldsymbol{\beta}} \mathrm{V}_{\boldsymbol{\beta}} + \mathrm{V}_{\boldsymbol{\lambda}\boldsymbol{\beta}})\mathrm{S}^{-1}_{\boldsymbol{\beta}} & 
\mathrm{S}^{-1}_{\boldsymbol{\lambda}}(\mathrm{L} + \mathrm{V}_{\boldsymbol{\lambda}})\mathrm{S}^{-1}_{\boldsymbol{\lambda}}
\end{pmatrix},
\end{equation*}
where $\mathrm{L} = \mathrm{S}_{\boldsymbol{\lambda}\boldsymbol{\beta}}\mathrm{S}^{-1}_{\boldsymbol{\beta}}(\mathrm{V}_{\boldsymbol{\beta}} \mathrm{S}^{-1}_{\boldsymbol{\beta}}\mathrm{S}^{\top}_{\boldsymbol{\lambda}\boldsymbol{\beta}} - \mathrm{V}^{\top}_{\boldsymbol{\lambda}\boldsymbol{\beta}}) - \mathrm{V}_{\boldsymbol{\lambda}\boldsymbol{\beta}}\mathrm{S}^{-1}_{\boldsymbol{\beta}}\mathrm{S}^{\top}_{\boldsymbol{\lambda}\boldsymbol{\beta}}$. 

Moreover, note that $\mathrm{S}^{-1}_{\boldsymbol{\beta}}\mathrm{V}_{\boldsymbol{\beta}}\mathrm{S}^{-1}_{\boldsymbol{\beta}} = \mathrm{V}^{-1}_{\boldsymbol{\beta}}$, it shows that for known dispersion parameters, the asymptotic variance of the quasi-likelihood regression estimators 
reaches the Cramer Rao lower bound, which in turn shows that the quasi-likelihood approach provides asymptotically efficient estimators for the regression coefficients.

\citet{Jorgensen:2004} proposed the modified chaser algorithm to solve the system of equations $\mathcal{U}^{q}_{\boldsymbol{\beta}} = \boldsymbol{0}$ and $\mathcal{U}^{q}_{\boldsymbol{\lambda}} = \boldsymbol{0}$, defined by
\begin{eqnarray*}
\label{chaser}
\boldsymbol{\beta}^{(i+1)} &=& \boldsymbol{\beta}^{(i)} - \mathrm{S}_{\boldsymbol{\beta}}^{-1} \mathcal{U}^q_{\boldsymbol{\beta}}(\boldsymbol{\beta}^{(i)}, \boldsymbol{\lambda}^{(i)}) \nonumber \\
\boldsymbol{\lambda}^{(i+1)} &=& \boldsymbol{\lambda}^{(i)} - \mathrm{S}_{\boldsymbol{\lambda}}^{-1} \mathcal{U}^q_{\boldsymbol{\lambda}}(\boldsymbol{\beta}^{(i+1)}, \boldsymbol{\lambda}^{(i)}).
\end{eqnarray*}
The modified chaser algorithm uses the insensitivity property (\ref{Sbetalambda}), which allows us to use two
separate equations to update $\boldsymbol{\beta}$ and $\boldsymbol{\lambda}$.

\subsection{Pseudo-likelihood estimation}

We shall now present the pseudo-likelihood approach using terminology and results
from \citet{Gourieroux:1984}. The pseudo-likelihood approach considers the properties 
of estimators obtained by maximizing a likelihood function associated with a 
family of probability distributions, which does not necessarily contain the true 
distribution. In particular, in this paper to estimation of Tweedie regression
models, we adopted the Gaussian pseudo-likelihood, whose logarithm is given by 

\begin{equation}
\label{plik}
\mathcal{L}^p(\boldsymbol{\theta}) =-\frac{n}{2} \log(2\pi) - \frac{n \delta}{2} - \frac{p}{2}\sum_{i=1}^n \left( \log 
\mu_{i} - \frac{(y_{i}-\mu_{i})^2}{2 \exp(\delta) \mu_{i}^p} \right).
\end{equation}

The pseudo-score function for $\boldsymbol{\theta}$ is given by

\begin{equation*}
\mathcal{U}^{p}_{\boldsymbol{\theta}}(\boldsymbol{\beta}, \boldsymbol{\lambda}) = \left ( \frac{\partial \mathcal{L}^p(\boldsymbol{\theta})}{\partial \beta_0}^\top, \ldots, \frac{\partial \mathcal{L}^p(\boldsymbol{\theta})}{\partial \beta_Q}^{\top}, 
\frac{\partial \mathcal{L}^p(\boldsymbol{\theta})}{\partial \delta}^\top, \frac{\partial \mathcal{L}^p(\boldsymbol{\theta})}{\partial p}^\top \right )^\top,
\end{equation*}
whose components have the following form
\begin{equation}
\label{pseudoScore}
\frac{\partial \mathcal{L}^p(\boldsymbol{\theta})}{\partial \beta_j} = -\frac{p}{2} \sum_{i=1}^n x_{ij} + \sum_{i=1}^n\frac{p(y_{i}-\mu_{i})^2}{2\exp(\delta) \mu_{i}^p} x_{ij} + \sum_{i=1}^n \frac{(y_{i}-\mu_{i})}{\exp(\delta) \mu_{i}^{p-1}} x_{ij},
\end{equation}
\begin{equation}
\label{pseudophi}
\frac{\partial \mathcal{L}^p(\boldsymbol{\theta})}{\partial \delta} = -\frac{n}{2} + \frac{1}{2 \exp(\delta)} \sum_{i=1}^n \frac{(y_{i}-\mu_{i})^2}{\mu_{i}^p}
\end{equation}
and
\begin{equation}
\label{pseudop}
\frac{\partial \mathcal{L}^p(\boldsymbol{\theta})}{\partial p} = -\frac{1}{2}\sum_{i=1}^n \log(\mu_{i}) + \frac{1}{2 \exp(\delta)} \sum_{i=1}^n\frac{\log(\mu_{i})}{\mu_{i}^p}(y_{i}-\mu_{i})^2.
\end{equation}
We note in passing that Equation (\ref{pseudoScore}) is an unbiased estimating function for $\beta_j$ based on the linear and squared residuals. Similarly, note that Equations (\ref{pseudophi}) and (\ref{pseudop}) are unbiased estimating functions for $\delta$ and $p$
based on the squared residuals. 

\citet{Gourieroux:1984} showed under classical assumptions, that the pseudo-likelihood estimators denoted by $\boldsymbol{\hat{\theta}_{PL}}$ and obtained by maximizing Equation (\ref{plik}) converge almost surely to $\boldsymbol{\theta}$. Furthermore, $\boldsymbol{\hat{\theta}_{PL}}$ converges in distribution to $\mathrm{N}(\boldsymbol{\theta}, \mathcal{S}^{-1}_{\boldsymbol{\theta}} \mathcal{V}_{\boldsymbol{\theta}} \mathcal{S}^{-1}_{\boldsymbol{\theta}})$ where

\begin{equation*}
\mathcal{S}_{\boldsymbol{\theta}} = \mathrm{E} \left(- \frac{\partial^2 \mathcal{L}^p(\boldsymbol{\theta})}{\partial \boldsymbol{\theta} \partial \boldsymbol{\theta}^{\top}}    \right) \quad \text{and} \quad \mathcal{V}_{\boldsymbol{\theta}} = \mathrm{E}\left( \mathcal{U}^{p}_{\boldsymbol{\theta}}(\boldsymbol{\beta}, \boldsymbol{\lambda}) \mathcal{U}^{p}_{\boldsymbol{\theta}}(\boldsymbol{\beta}, \boldsymbol{\lambda})^{\top} \right ).
\end{equation*}
Similarly, the variability matrix (\ref{VTHETA}) in the context of quasi-likelihood estimation, the matrix $\mathcal{V}_{\boldsymbol{\theta}}$ depends on third and fourth moments. Hence, we propose to use the empirical version of $\mathcal{V}_{\boldsymbol{\theta}}$, which is given by
\begin{equation*}
\tilde{\mathcal{V}}_{\boldsymbol{\theta}} = \sum_{i=1}^n \mathcal{U}^{p}_{\boldsymbol{\theta}}(\boldsymbol{\theta})_i \mathcal{U}^{p}_{\boldsymbol{\theta}}(\boldsymbol{\theta})_i,
\end{equation*}
where the sum is understood to be element-wise. We shall now compute the components of the $\mathcal{S}_{\boldsymbol{\theta}}$. First, note that the matrix $\mathcal{S}_{\boldsymbol{\theta}}$ can be partitioned as

\begin{equation*}
\label{Vtheta}
\mathcal{S}_{\boldsymbol{\theta}} = \begin{pmatrix}
\mathcal{S}_{\boldsymbol{\beta}} & \mathcal{S}_{\boldsymbol{\beta} \delta} & \mathcal{S}_{\boldsymbol{\beta}p}  \\ 
\mathcal{S}_{\delta \boldsymbol{\beta}} & \mathcal{S}_{\delta} & \mathcal{S}_{\phi p} \\
\mathcal{S}_{p \boldsymbol{\beta}}    & \mathcal{S}_{p \phi} & \mathcal{S}_{p} \\
\end{pmatrix}.
\end{equation*}

The entry $(j,k)$ of the $Q \times Q$ matrix $\mathcal{S}_{\boldsymbol{\beta}}$ is given by

\begin{equation*}
\mathcal{S}_{\beta_{jk}} = \sum_{i=1}^n \left( \frac{p^2 x_{ij} x_{ik}}{2} + \frac{x_{ij}x_{ik}}{\exp(\delta) \mu_i^{p-2}}\right).
\end{equation*}

Similarly, the entries $\mathcal{S}_{\delta}$ and $\mathcal{S}_{p}$ are respectively given by

\begin{equation*}
\mathcal{S}_{\delta} = \frac{n}{2} \quad \text{and} \quad \mathcal{S}_{p} = \sum_{i=1}^n\frac{\log(\mu_i)^2}{2}.
\end{equation*}

Furthermore, the cross entries have the form

\begin{equation*}
\mathcal{S}_{\beta_j \delta} = \sum_{i=1}^n \frac{p x_{ij}}{2}, \quad \mathcal{S}_{\beta_j p} = \sum_{i=1}^n\frac{log(\mu_i)x_{ij}-p}{2} \quad \text{and} \quad \mathcal{S}_{\delta p} = \sum_{i=1}^n \frac{\log(\mu_i)}{2}.
\end{equation*}

Finally, we propose the Newton scoring algorithm to solve the system of equations $\mathcal{U}^{p}_{\boldsymbol{\theta}}(\boldsymbol{\beta}, \boldsymbol{\lambda}) = \boldsymbol{0}$, defined by
\begin{eqnarray}
\label{chaser}
\boldsymbol{\theta}^{(i+1)} &=& \boldsymbol{\theta}^{(i)} - \mathcal{S}_{\boldsymbol{\theta}}^{-1} \mathcal{U}^p_{\boldsymbol{\theta}}(\boldsymbol{\beta}^{(i)}, \boldsymbol{\lambda}^{(i)}). \nonumber
\end{eqnarray}
In that case, we have to update $\boldsymbol{\beta}$ and $\boldsymbol{\lambda}$ together, since the cross-entries of $\mathcal{S}_{\boldsymbol{\theta}}$ are not zeroes.

\section{Simulation studies} \label{simulation}
In this section we shall present two simulation studies designed to
i) check the asymptotic properties of the maximum, quasi- and 
pseudo-likelihood estimators in a finite sample scenario and ii) check the
robustness of the Tweedie regression models in the case of 
misspecification by heavy tailed distributions.
    \subsection{Fitting Tweedie regression models}\label{simulation1}
In this section we present a simulation study that was conducted to
compare the properties of the estimation methods. We evaluated the 
expected bias, consistency, coverage rate and efficiency for the maximum 
likelihood (MLE), quasi-likelihood (QMLE) and pseudo-likelihood (PMLE) 
estimators. We generated $1000$ data sets considering four sample sizes 
$100, 250, 500$ and $1000$. We considered five values of the power 
parameter $0, 1.01, 1.5, 2$ and $3$ combined with three amounts of 
variation. We used the average coefficient of variation to measure the 
amount of variation introduced in the data. We defined, small, medium 
and large amount of variation data sets generated using coefficient
of variation equals to $15\%$, $50\%$ and $80\%$, respectively. 
The values of the power parameter were chosen to have 
non-standard situations, as the cases of $p=0$ and $p=1.01$ where 
we expect the MLE does not work. The case of $p=2$ is also difficult
for maximum likelihood estimation, since the probability density 
function should be evaluated using two different infinity sums, for
$p < 2$ and $p > 2$. The cases $p=1.5$ and $p=3$ represent 
the standard compound Poisson and inverse Gaussian distributions, 
respectively. In these cases, we expect that the MLE works well, so
we have safe results to compare with our two alternative approaches. 

\begin{figure}
\centering
\includegraphics[angle=90]{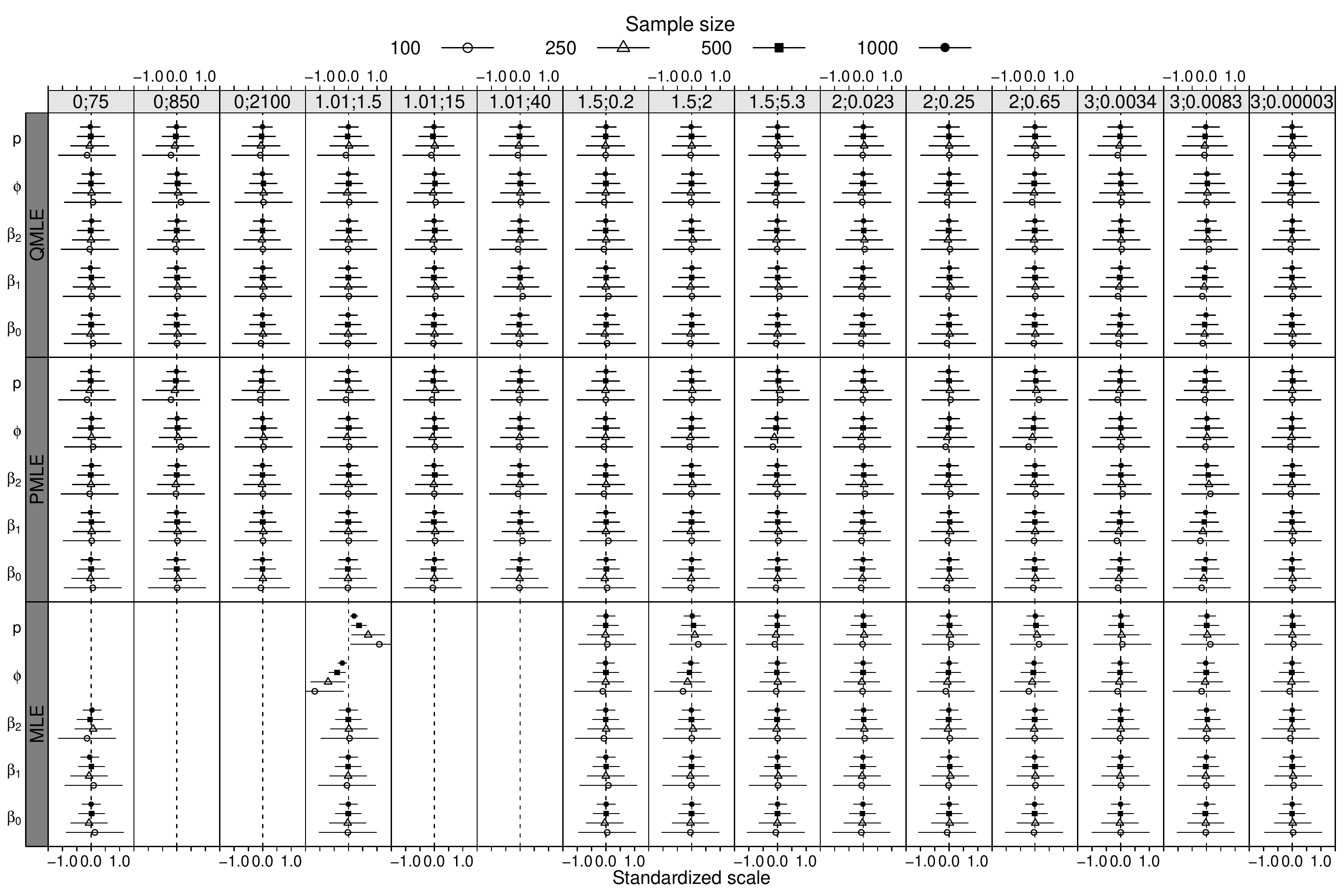}
\caption{Expected bias and confidence interval on a standardized scale by
estimation methods (maximum likelihood (MLE), pseudo-likelihood (PMLE) and quasi-likelihood (QMLE)),
sample size and different values of the power and dispersion parameters ($p;\phi$).}
\label{fig:simul}
\end{figure}

All scenarios consider models with an intercept ($\beta_0 = 2$) and 
slopes ($\beta_1 = 0.8$, $\beta_2 = -1.5$). The covariates are a sequence 
from $-1$ to $1$, representing a continuous covariate, a factor with 
two levels ($0$ and $1$) and length equals the sample size.
For $p=0$ the dispersion parameter values are $\phi = (75, 850, 2100)$
corresponding, respectively, to small ($15\%$), medium ($50\%$) and
large ($80\%$) variation. 
Similarly, for $p=1.01$, $p=1.5$, $p=2$ and $p=3$
the dispersion parameter values are $\phi = (1.5, 15, 40)$, 
$\phi = (0.2, 2, 5.3)$, $\phi = (0.023, 0.25, 0.65)$ and 
$\phi = (0.0003, 0.0034, 0.0083)$, respectively. 
Fig.~\ref{fig:simul} shows the expected bias plus and minus the expected 
standard error for the parameters on each model and scenario. 
The scales are standardized for each parameter dividing the expected bias 
and the limits of the confidence intervals by the standard error obtained 
on the sample of size $100$.

\setkeys{Gin}{width=500pt}

\begin{figure}
\centering
\includegraphics{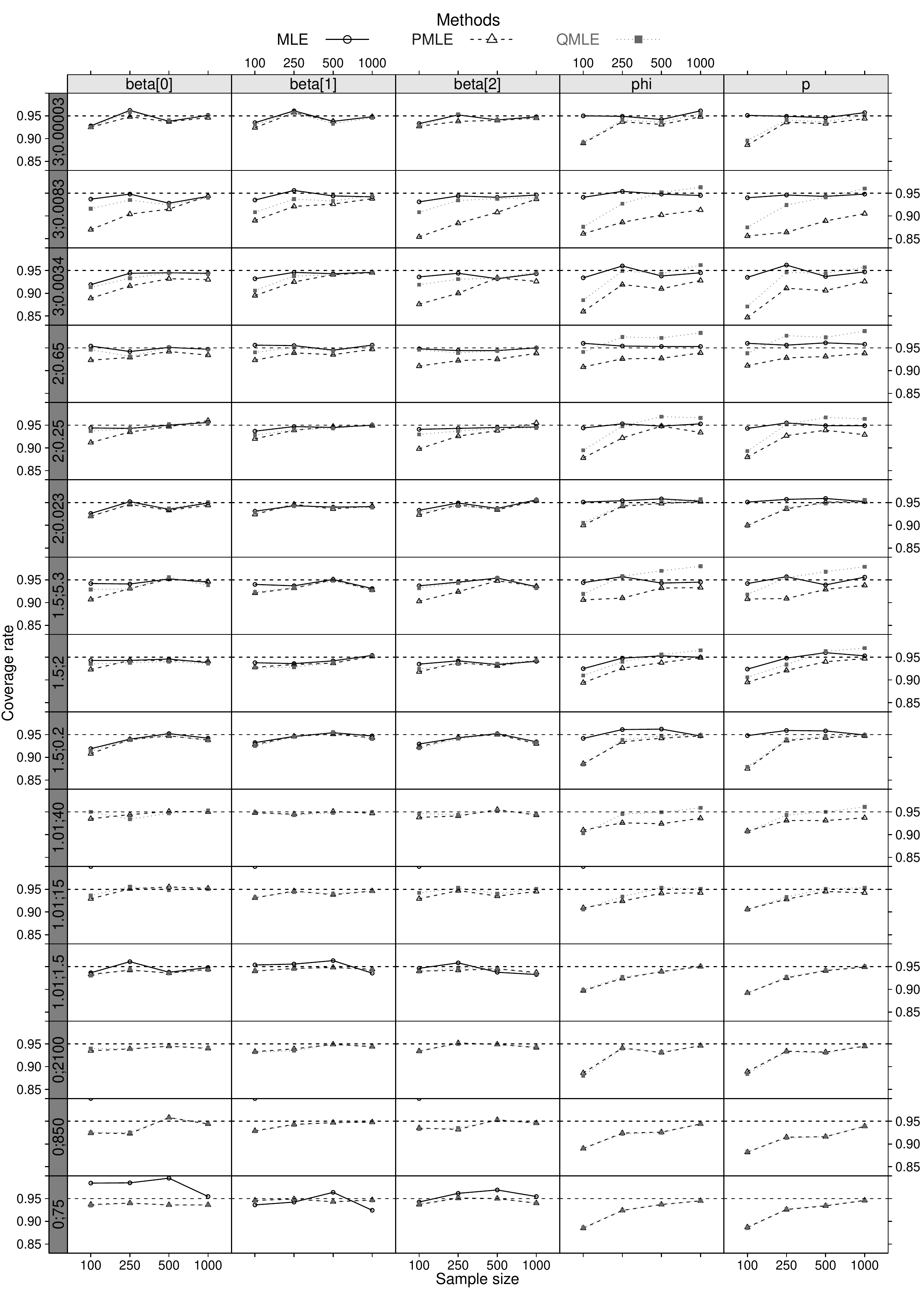}
\caption{Coverage rate for each parameter ($\beta_0,\beta_1,\beta_2,\phi,p$) 
by estimation methods (maximum likelihood (MLE), 
pseudo-likelihood (PMLE) and quasi-likelihood (QMLE)), sample size
and different values of the power and dispersion parameters ($p;\phi$).}
\label{fig:coverage}
\end{figure}

The results in Fig.~\ref{fig:simul} show that for the quasi- and
pseudo-likelihood methods and all simulation scenarios, both the expected 
bias and standard error tend to $0$ as the sample size is increased.
It shows the consistency and unbiasedness of our estimators.
As expected the maximum likelihood method did not work for $p=0$ and
$p=1.01$ in the medium and large variation scenarios. In these cases,
the algorithm failed for all simulated data sets.
In the cases of small variation the algorithm converged for 
$132$ and $326$ data sets for $p = 0$ and $p=1.01$, respectively. 
In these scenarios, although the large bias for the dispersion 
parameters, the regression coefficients were consistently estimated.
Fig.~\ref{fig:coverage} presents the coverage rate by estimation
methods, sample size and simulation scenarios.

\setkeys{Gin}{width=370pt}

\begin{figure}
\centering
\includegraphics{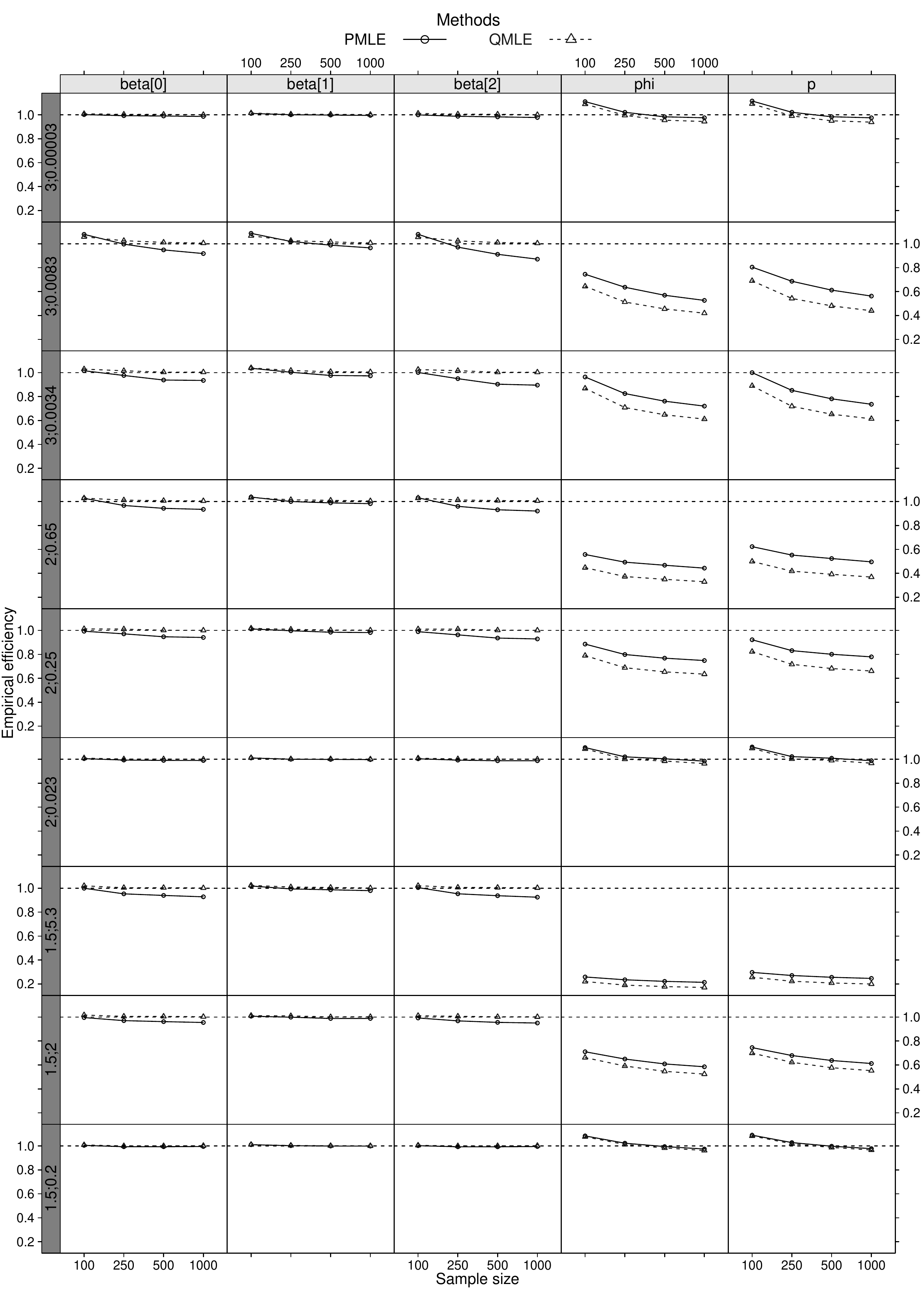}
\caption{Empirical efficiency for each parameter ($\beta_0,\beta_1,\beta_2,\phi,p$) 
by estimation methods (maximum likelihood (MLE), 
pseudo-likelihood (PMLE) and quasi-likelihood (QMLE)), sample size
and different values of the power and dispersion parameters ($p;\phi$).}
\label{fig:efficiency}
\end{figure}

The results presented in Fig.~\ref{fig:coverage} show that in general 
for large samples the coverage rates are close to the nominal level~$(0.95)$ 
for all parameters and simulation scenarios. The MLE presented coverage rate
zero for the dispersion parameters, when $p = 0$ and $p=1.01$ in all 
simulation scenarios~(not shown). 
The quasi-likelihood method presented coverage rate closer to the nominal 
level than the pseudo-likelihood method, mainly for dispersion parameters
and large values of the power parameter $(p \geq 1.5)$.
Regarding the estimation methods as expected the MLE
presented the coverage rate close to the nominal level for large 
values of the power parameter. The alternative approaches worked well in all 
simulation scenarios, including the cases where the MLE did not work.
Finally, Fig.~\ref{fig:efficiency} presents the empirical efficiency of 
the quasi- and pseudo-likelihood estimators. The empirical efficiency was 
computed as the ratio between the variance of the MLE and the variance
obtained by the alternative approaches. We computed the efficiency only
for the cases where $p \geq 1.5$, since for the other cases the MLE 
presented no reliable results.

The results in Fig.\ref{fig:efficiency} show that for the regression coefficients
both QMLE and PMLE approaches presented efficiency close to $1$ in all
simulation scenarios. Concerns the dispersion parameters, for the small
variation scenario the QMLE and PMLE presented efficiency 
close to $1$. However, when the variation increased these estimators 
loss efficiency, the worst scenario appears for $p=1.5$ and large
variation, where the efficiency presented values around $20\%$.
In general the PMLE is more efficient than the QMLE for the dispersion
and power parameters.

\setkeys{Gin}{width=370pt}

\subsection{Robustness of Tweedie regression models}\label{simulation2}

In this subsection we present a simulation study that was conducted to
evaluate the robustness of the Tweedie regression models in the case of
model misspecification by heavy tailed distributions. 
We generated $1000$ data sets considering four sample sizes 
$100, 250, 500$ and $1000$ following two heavy tailed distributions, 
namely, t-Student and slash. The parametrization adopted was the one
implemented in the \texttt{R} package \texttt{heavy}~\citep{heavy:2016}.
For both distributions, we designed three simulation scenarios according 
to the amount of variation introduced in the data. 
We defined, small, medium and large amount of variation data sets 
generated using dispersion parameter equals to $100$, $500$ and $1000$, 
respectively. In order to simulate challenge data sets, we used 
$2$ degrees of freedom.
The mean structure was specified as in the 
subsection~\ref{simulation1}. In the case of heavy tailed distributions, 
we expect negative values for the power parameter. 
Thus, we fitted the Tweedie regression models by using the quasi- and 
pseudo-likelihood approaches. 

\begin{figure}
\centering
\includegraphics{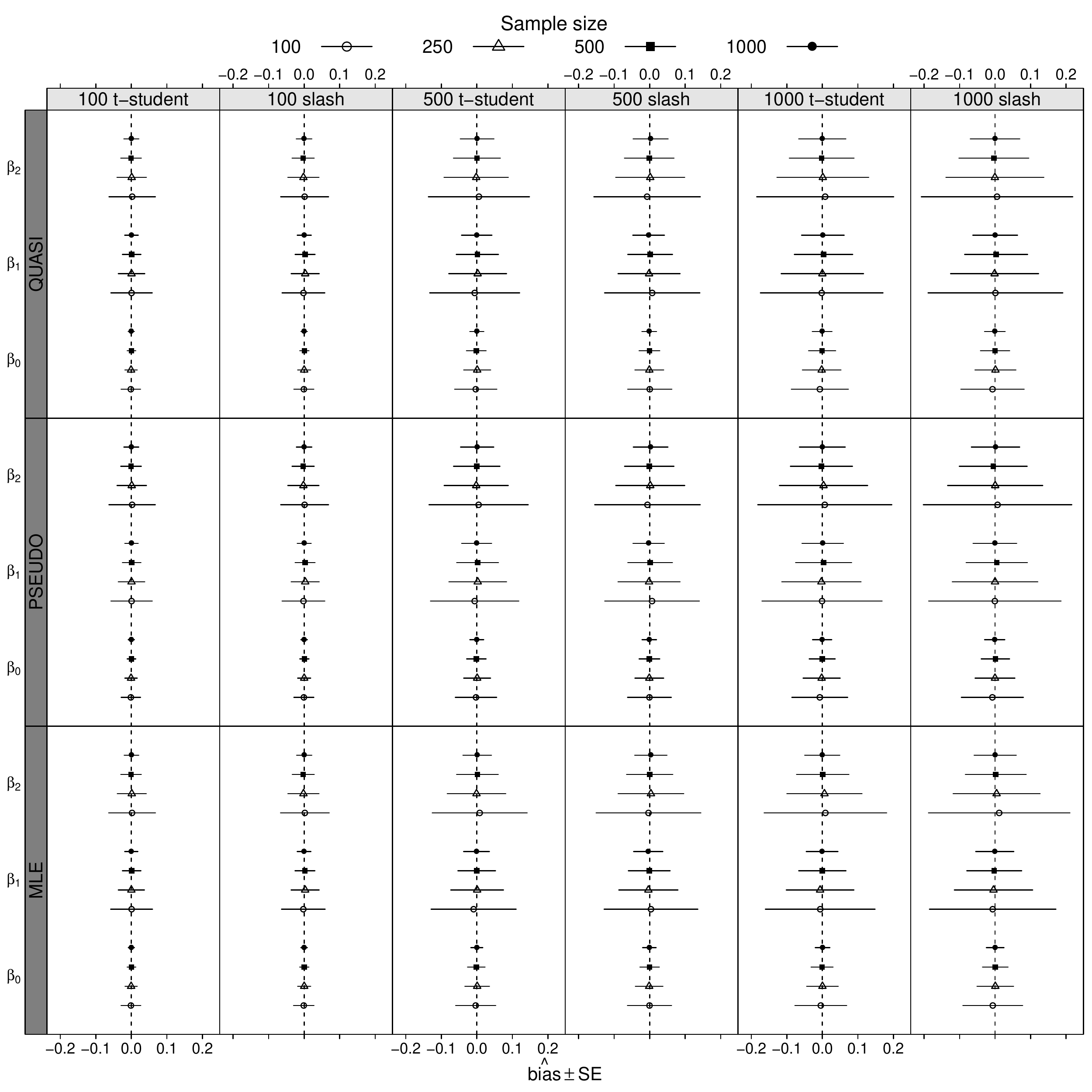}
\caption{Expected bias and confidence interval by estimation methods 
(quasi-likelihood (QMLE), pseudo-likelihood (PMLE) and maximum likelihood (MLE)),
sample size and simulation scenarios.}
\label{fig:simul2}
\end{figure}

In order to compute the empirical efficiency of the quasi- and 
pseudo-likelihood estimators, we fitted t-Student regression models
along with the logarithm link function, as implemented in the package
\texttt{gamlss}(family \texttt{TF})~\citep{Rigby:2005}. 
Although, of the extensive literature on robust estimation methods, 
in this paper we adopted the t-Student regression models, since it is
a frequent choice for the analysis of heavy tailed data~\citep{robust:2009}
and can be fitted using the orthodox maximum likelihood method.
Furthermore, since there is no software available for fitting slash 
regression models using logarithm link function, the t-Student 
regression models were used as the base of comparison for both
t-Student and slash data sets. Fig.~\ref{fig:simul2} shows the expected 
bias plus and minus the expected standard error for the regression 
parameters by estimation methods, sample size and simulation scenarios.

\setkeys{Gin}{width=0.99\textwidth}

\begin{figure}
\centering
\includegraphics{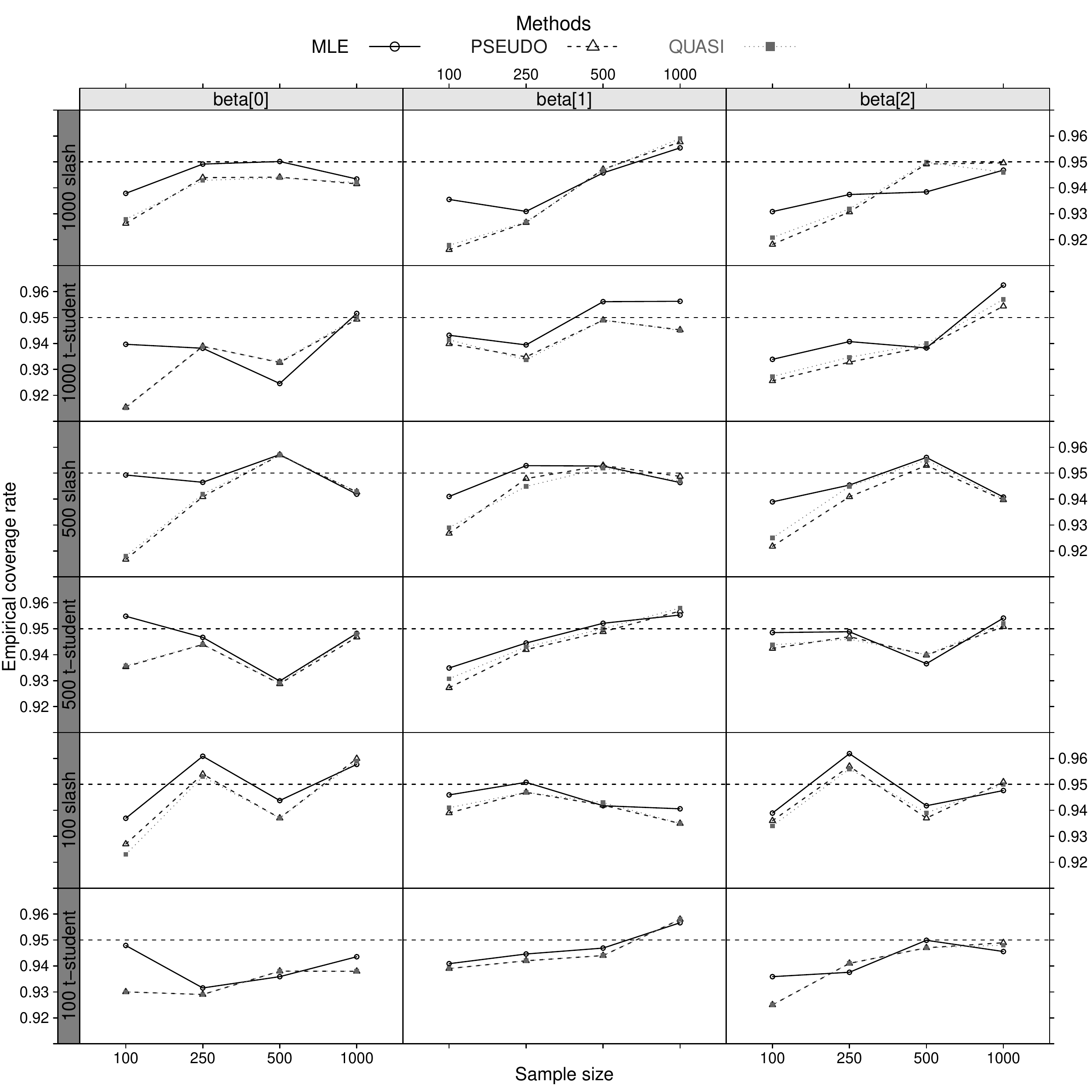}
\caption{Coverage rate for regression parameters by estimation methods 
(quasi-likelihood (QMLE), pseudo-likelihood (PMLE) and maximum likelihood (MLE)), 
sample size and simulation scenarios.}
\label{fig:coverage2}
\end{figure}

The results presented in Fig.~\ref{fig:simul2} show that the three 
estimation methods provide unbiased and consistent estimates of the 
regression parameters in all simulation scenarios. As expected, the
standard errors associated with the regression parameters increase while
the amount of variation introduced in the data increases.
Fig.~\ref{fig:coverage2} presents the coverage rate by estimation
methods, sample size and simulation scenarios.

\setkeys{Gin}{width=0.99\textwidth}

\begin{figure}
\centering
\includegraphics{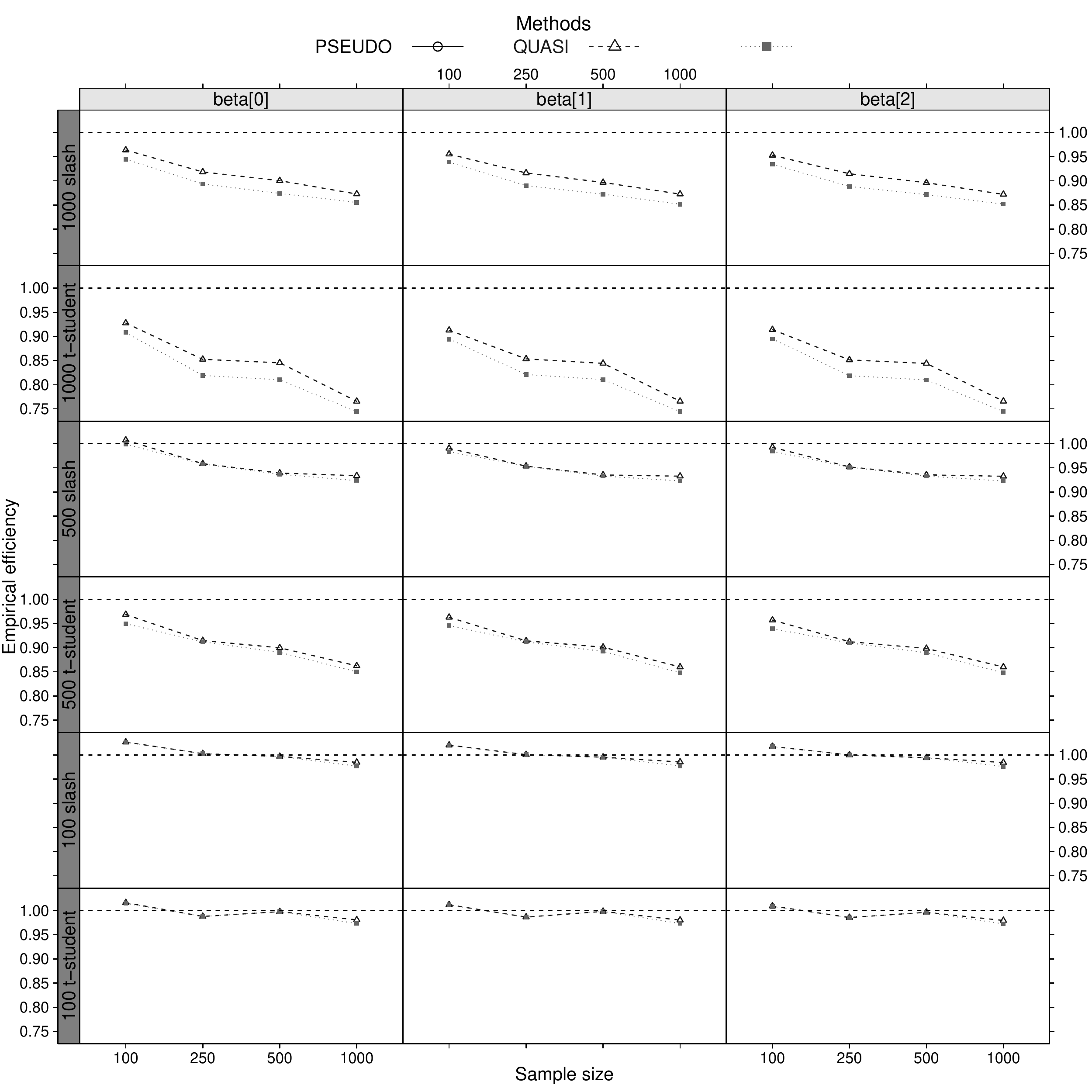}
\caption{Empirical efficiency for regression parameters by estimation 
methods (quasi-likelihood (QMLE), pseudo-likelihood (PMLE) and maximum likelihood (MLE)), 
sample size and simulation scenarios.}
\label{fig:efficiency2}
\end{figure}

The empirical coverage rate presented values close to the nominal specified 
level of $95\%$ for all estimation methods and simulation scenarios.
The MLE method presented coverage rate closer to the nominal level than
the QMLE and PMLE methods, however, the difference is no larger than $3\%$.
The coverage rate of the QMLE and PMLE were virtually the same for
all regression parameters, sample size and simulation scenarios.
Finally, Fig.~\ref{fig:efficiency2} presents the empirical efficiency of 
the QMLE and PMLE estimators for the regression parameters. The empirical
efficiency was computed as the ratio between the variance of the MLE 
obtained by fitting the t-Student regression models and the variance
of the QMLE and PMLE estimators obtained by fitting the Tweedie regression
models.

\setkeys{Gin}{width=0.99\textwidth}

The empirical efficiency presented values close to $1$ for the small 
variation simulation scenarios, however, when the amount of variation 
increases both QMLE and PMLE loss efficiency. The loss were around $10\%$
and $20\%$ for the medium and large variation scenarios, respectively.
The results are worse for large samples. The PMLE presents efficiency
slightly closer to the nominal level than the QMLE.

\section{Data analyses}
In this section we shall present three illustrative examples of Tweedie regression models.
The data that are analysed and the programs that were used to analyse them
can be obtained from:\\ http://www.leg.ufpr.br/doku.php/publications:papercompanions:tweediereg.

\subsection{Smoothing time series of rainfall in Curitiba, Paran\'a, Brazil}

\begin{figure}
\centering
\includegraphics{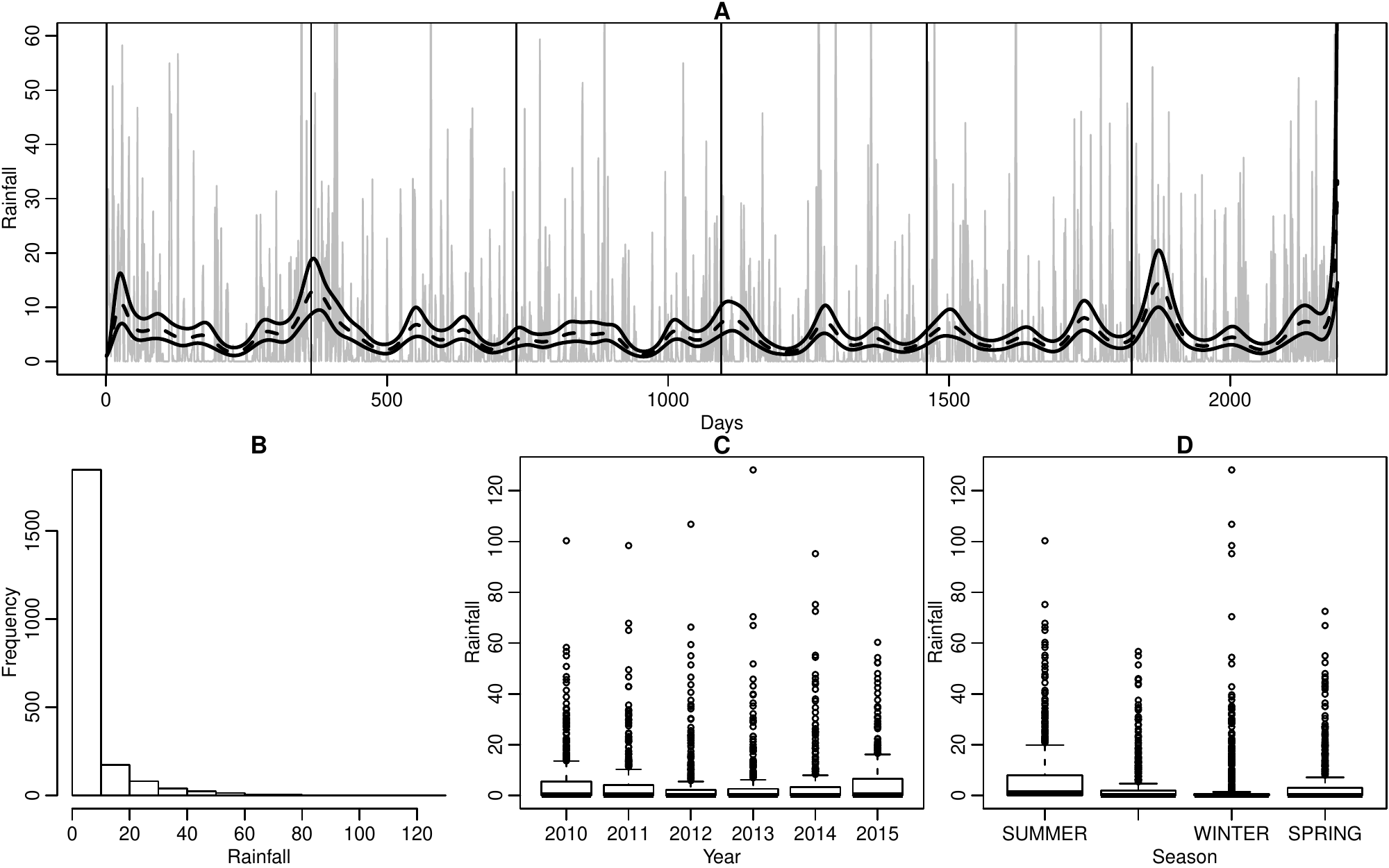}
\caption{Time series plot for Curitiba rainfall data with fitted values~(A).
Vertical black lines indicate January 1st.
Histogram of daily rainfall for the whole period~(B). 
Boxplots for year~(C) and season~(D).}
\label{fig:exemplo1}
\end{figure}

This example concerns daily rainfall data in Curitiba, Paran\'a State, 
Brazil. The data were collected for the period from $2010$ to $2015$ 
corresponding to $2191$ days. The main goal is to smooth the time 
series to help us better see patterns or trends. The analysis of rainfall
data is in general challenged by the presence of many zeroes and the 
highly right-skewed distribution of the data. The plots shown in 
Fig.~\ref{fig:exemplo1} illustrate some of these features for the
Curitiba rainfall data. In particular, Fig.~\ref{fig:exemplo1}(B) 
highlights the right-skewed distribution and the considerable proportion
of exact 0s ($51\%)$.

\setkeys{Gin}{width=0.9\textwidth}

In order to smooth the Curitiba rainfall time series, we fitted a Tweedie 
regression model with linear predictor expressed in terms of 
B-splines~\citep{Boor:1972}. The natural basis regression smoothing 
framework was used to select the degree of smoothness~\citep{Wood:2006}. 
In that case, we found that $14$ degrees of freedom were enough to 
smooth the times series. The models were fitted by using the three 
estimation methods, namely, maximum likelihood~(MLE), quasi-(QMLE) and 
pseudo-likelihood~(PMLE). Table~\ref{tab:disp1} presents estimates and 
standard errors for the dispersion and power parameters.

\begin{table}[h]
\centering
\caption{Dispersion and power parameter estimates and standard errors~(SE) 
by estimation methods for the Curitiba rainfall data.}
\label{tab:disp1}
\begin{tabular}{lcccccc} \hline
\multirow{3}{*}{Parameter} & \multicolumn{6}{c}{Estimation methods}        \\ \hline
         & \multicolumn{2}{c}{MLE} & \multicolumn{2}{c}{QMLE} &\multicolumn{2}{c}{PMLE}    \\ \hline
         & Estimate & SE           &  Estimate & SE           & Estimate & SE \\ \hline
$\delta$ & $2.0284$ & $0.0292$     & $2.2791$  & $0.2194$     & $2.8543$ & $0.4355$    \\
$p$      & $1.6774$ & $0.0089$     & $1.4721$  & $0.1455$     & $1.2652$ & $0.2492$ \\ \hline 
\end{tabular}
\end{table}

\begin{figure}
\centering
\includegraphics{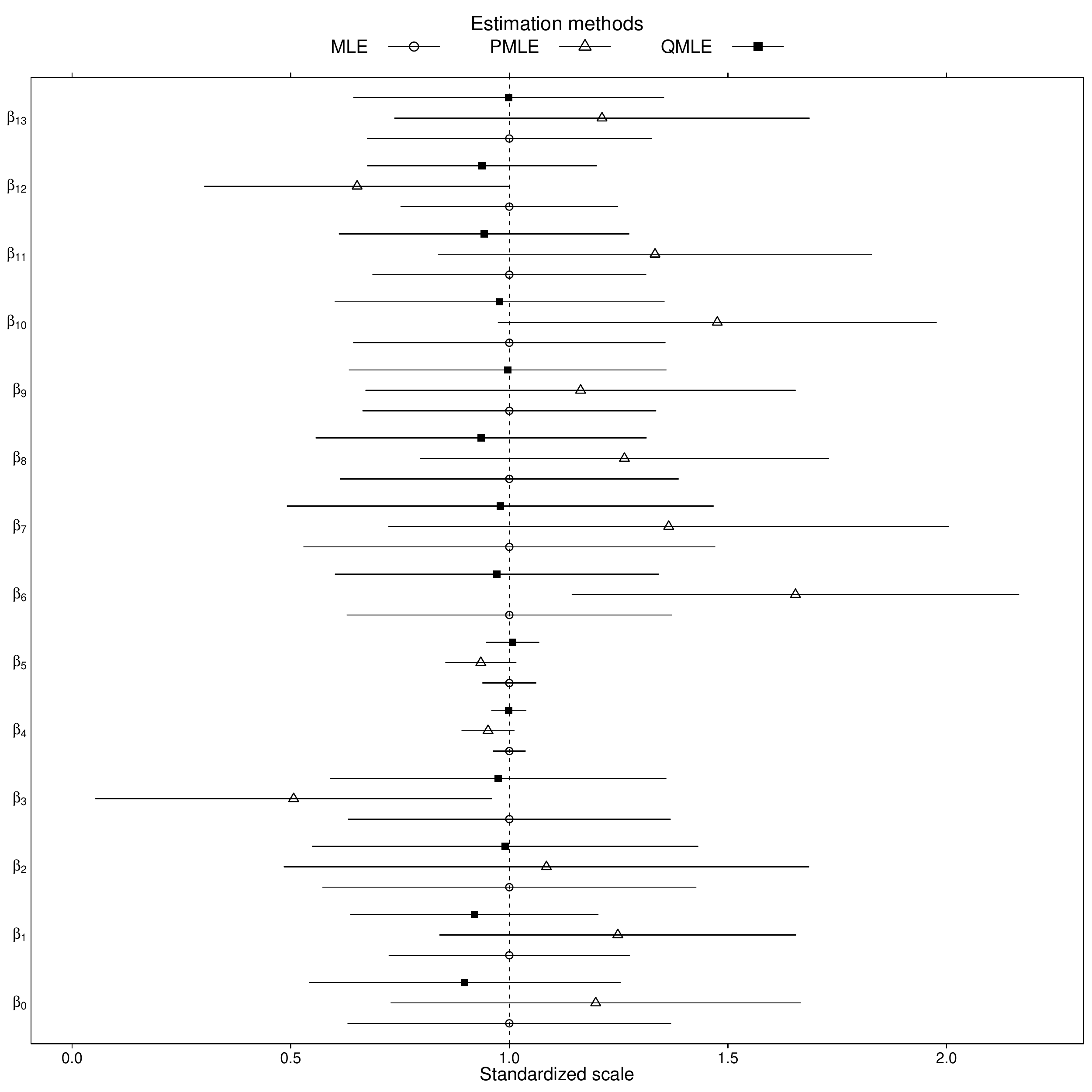}
\caption{Regression parameter estimates and $95\%$ confidence intervals by 
estimation methods for the Curitiba rainfall data.}
\label{fig:beta1}
\end{figure}

The results in Table~\ref{tab:disp1} show slightly different estimates
for the dispersion and power parameters, depending on the estimation
method used. However, the confidence intervals obtained by the QMLE and PMLE 
approaches contain the MLE. 
The standard errors obtained by the alternative approaches are larger 
than the ones obtained by the MLE. 
To evaluate the effect of the estimation methods on the regression 
coefficients, Fig.~\ref{fig:beta1} shows estimates and
confidence intervals for each regression coefficient by
estimation methods. The scales were standardized for each parameter 
dividing the estimate and the limits of the confidence interval by the 
estimate obtained by the maximum likelihood method.

\setkeys{Gin}{width=0.99\textwidth}

The results in Fig.~\ref{fig:beta1} show that the QMLE method
presented estimates and confidence intervals more similar to the MLE than 
the PMLE method. The relative average difference between the MLE and QMLE 
estimates was $3.36\%$. On the other hand, the relative average difference 
between the MLE and PMLE estimates was $14.58\%$.
Similarly, the confidence intervals obtained by the QMLE method were on 
average $3.33\%$ wider than the corresponding MLE intervals. 
On the other hand, the confidence intervals obtained by the 
PMLE approach were $39.98\%$ wider than the MLE intervals. 

For all estimation methods, the power parameter estimates are in the 
interval $1 < p < 2$, suggesting a compound Poisson distribution, 
as expected, since the response variable is continuous with exact 0s. 
The fitted values and $95\%$ confidence interval obtained by the 
quasi-likelihood method are shown in Fig.~\ref{fig:exemplo1} above.
The fitted values obtained by the MLE and PMLE approaches were
similar the ones obtained by the QMLE (not shown).
The smooth function captures the swing in the data and highlights 
the seasonal behaviour with dry and wet months around the winter and 
summer seasons, respectively.

In order to compare the computational times required by each approach 
for fitting the Tweedie regression model for this data set, 
we used the package \texttt{rbenchmark}~\citep{bench:2012}.
The computations were done by a standard personal computer at $2.90$ GHz
with $8$ G RAM by using the \texttt{R} software version $3.2.2$ for 
ten replications. The results showed that the QMLE approach is $37$ and 
$0.22$ times faster than the MLE and PMLE approaches, respectively.

\subsection{Income dynamics in Australia}
We consider some aspects of a cross-section study on earnings of $595$ 
individuals for the year $1982$ in Australia. 
The data set is available in the package \texttt{AER}~\citep{Klieber:2008} 
for the statistical software \texttt{R}. The response variable 
$\texttt{wage}$ is known to be highly-right skewed.
The data set has $12$ covariates: \texttt{experience} years of full-time
work experience; \texttt{weeks} weeks worked; \texttt{occupation} factor
two levels (white-collar, blue-collar); \texttt{industry} factor
two levels (no;yes) indicating if the individual work in a manufacturing
industry; \texttt{south} factor two levels (no;yes) indicating if the
individuals resides in the south; \texttt{smsa} factor two levels 
(no;yes) indicating if the individual resides in a standard metropolitan
area; \texttt{gender} factor indicating gender (male, female);
\texttt{union} factor two levels (no, yes) indicating if the individual's
wage set by a union contract; \texttt{ethnicity} factor indicating
ethnicity, African-American (afam) or not (other). 
The main goal of the investigation was to assess the effect of the 
covariates on the wage.
We fitted the Tweedie regression model with linear predictor composed
by all covariates by using the three estimation methods. 
Table~\ref{tab:disp2} shows the estimates and standard errors for the 
regression, dispersion and power parameters.

\begin{table}[h]
\centering
\caption{Regression, dispersion and power parameter estimates and standard errors~(SE) 
by estimation methods for the income dynamics data.}
\label{tab:disp2}
\begin{tabular}{lcccccc} \hline
\multirow{3}{*}{Parameter} & \multicolumn{6}{c}{Estimation methods}        \\ \hline
            & \multicolumn{2}{c}{MLE} & \multicolumn{2}{c}{QMLE} &\multicolumn{2}{c}{PMLE}    \\ \hline
                    & Estimate & SE   & Estimate & SE   & Estimate & SE \\  \hline
\texttt{Intercept}  & 5.8580 & 0.1723 & 5.8480 & 0.1813 & 5.9137 & 0.1859 \\ 
\texttt{experience}  & 0.0056 & 0.0013 & 0.0056 & 0.0014 & 0.0068 & 0.0013 \\ 
\texttt{weeks}      & 0.0034 & 0.0026 & 0.0035 & 0.0028 & 0.0041 & 0.0030 \\ 
\texttt{occupation} & -0.1870 & 0.0365 & -0.1893 & 0.0362 & -0.1977 & 0.0352 \\ 
\texttt{industry}   & 0.0716 & 0.0293 & 0.0731 & 0.0302 & 0.0229 & 0.0322 \\ 
\texttt{south}      & -0.0375 & 0.0305 & -0.0363 & 0.0320 & -0.0104 & 0.0341 \\ 
\texttt{smsa}       & 0.1644 & 0.0293 & 0.1658 & 0.0297 & 0.1456 & 0.0312 \\ 
\texttt{married}    & 0.1172 & 0.0478 & 0.1218 & 0.0523 & 0.0902 & 0.0538 \\ 
\texttt{gender}     & -0.3389 & 0.0570 & -0.3346 & 0.0567 & -0.4039 & 0.0562 \\ 
\texttt{union}      & 0.1265 & 0.0314 & 0.1331 & 0.0298 & 0.0839 & 0.0293 \\ 
\texttt{education}  & 0.0577 & 0.0065 & 0.0578 & 0.0069 & 0.0543 & 0.0074 \\ 
\texttt{ethnicity}  & -0.1793 & 0.0506 & -0.1772 & 0.0510 & -0.1466 & 0.0484 \\  \hline
$\delta$            & -5.9848 & 1.1117 & -6.8587 & 2.0409 & -7.1317 & 1.8857 \\ 
$p$                 & 2.5354 & 0.1605 & 2.6656 & 0.2979 & 2.7012 & 0.2735 \\ \hline
\end{tabular}
\end{table}

The results in Table~\ref{tab:disp2} show that the MLE
and QMLE approaches strongly agree in terms of estimates
and standard errors for the regression coefficients. 
The PMLE approach presents estimates slightly different from the MLE and QMLE 
approaches. Regarding the dispersion parameters, although the slightly
difference in terms of estimates and standard errors, the confidence intervals 
from the QMLE and PMLE approaches contain the MLE estimates. 

Concerning the effect of the covariates the MLE and QMLE approaches
agree that the covariates \texttt{weeks} and \texttt{south} are 
non-significant. On the other hand, the PMLE approach also indicated 
that the covariates \texttt{industry} and \texttt{married} are 
non-significant. Regarding the other covariates the three approaches
agree that they are significant.

In order to compare the fit of Tweedie regression model with more standard 
approaches, we also fitted the Gaussian, gamma and inverse Gaussian regression 
models for the income dynamics data set. The maximized values of the 
log-likelihood function were $-4437.51$, $-4318.08$ and $-4316.52$
for the Gaussian, gamma and inverse Gaussian models, respectively.
Furthermore, the maximized value of the log-likelihood function
for the Tweedie regression model was $-4312.39$, which in turn 
shows the better fit of the Tweedie regression model, as expected.
In terms of computational time for this data set, the QMLE 
approach was $45$ and $0.15$ times faster than the MLE and PMLE 
approaches, respectively.

\subsection{Gain in weight of rats}
The third example concerns to a standard Gaussian regression model.
The goal of this example is to show that the quasi- and pseudo-likelihood
approaches can estimate values of the power parameter between
$0$ and $1$, where the maximum likelihood estimator does not exist.
We used the \texttt{weightgain} data set available in the \texttt{HSAUR}
package~\citep{HSAUR:2015}. This data set corresponds to an experiment
to study the gain in weight of rats fed on four different diets, 
distinguished by the amount of protein (low and high) and by source of 
protein (beef and cereal). The data set has $40$ observations.

We fitted the Gaussian, gamma, inverse Gaussian and Tweedie regression
models for the \texttt{weightgain} data set. The linear predictor was
composed of the two main covariates \texttt{source} and \texttt{type} 
along with the interaction term, for all models.
The values of the maximized log-likelihood were $-162.84$, $-164.21$, 
$-165.36$ and $-163.50$ for the Gaussian, gamma, inverse Gaussian and
Tweedie models, respectively. These results showed that the Gaussian
distribution provides the best fit for this data set, judging by the
maximized log-likelihood value. In that case, the MLE method is not
able to indicate the best fit. It is due to the non-trivial restriction
on the power parameter space. Thus, we fitted the model using
the approaches QMLE and PMLE.
Table~\ref{tab:disp3} presents the estimates and standard errors for the 
regression, dispersion and power parameters, obtained by MLE, QMLE and
PMLE approaches.

\begin{table}[h]
\centering
\caption{Regression, dispersion and power parameter estimates and standard errors~(SE) 
by estimation methods for the gain in weight of rats data.}
\label{tab:disp3}
\begin{tabular}{lcccccc} \hline
\multirow{3}{*}{Parameter} & \multicolumn{6}{c}{Estimation methods}        \\ \hline
            & \multicolumn{2}{c}{MLE} & \multicolumn{2}{c}{QMLE} &\multicolumn{2}{c}{PMLE}    \\ \hline
                    & Estimate   & SE   & Estimate    & SE   & Estimate    & SE \\  \hline
\texttt{Intercept}  & $4.5891$   & 0.0504 & $4.6051$  & 0.0454 & $4.6050$  & 0.0453 \\ 
\texttt{source}     & $-0.1263$  & 0.0734 & $-0.1519$ & 0.0693 & $-0.1517$ & 0.06867 \\ 
\texttt{type}       & $-0.2235$  & 0.0750 & $-0.2331$ & 0.0694 & $-0.2337$ & 0.06922 \\ 
\texttt{source:type}& $0.1827$   & 0.1069 & $0.2096$  & 0.1036 & $0.2108$  & 0.1026 \\ \hline
$\delta$            & $0.6323$   & 8.1352 & $3.3614$  & 8.7203 & $3.3355$  & 9.0088 \\ 
$p$                 & $1.0590$   & 1.8400 & $0.4350$  & 1.9484 & $0.4408$  & 2.0129 \\ \hline
\end{tabular}
\end{table}

The results in Table~\ref{tab:disp3} show that the three approaches
strongly agree in terms of estimates and standard errors for the
regression coefficients. The value of the power parameter was estimated
smaller than $1$ by the QMLE and PMLE approaches, as expected, since 
the Gaussian distribution provides the best fit for this data. 
On the other hand, the maximum likelihood method estimated the power 
parameter close to $1$ the border of the parameter space, in that case
a non-optimum model. All approaches presented large standard errors 
for the power and dispersion parameters. In terms of computation time, 
for this application the PMLE approach was $94$ and $0.15$ times faster 
than the MLE and QMLE approaches, respectively.

\section{Discussion}

In this paper, we adopted the quasi- and pseudo-likelihood 
approaches to estimation and inference of Tweedie regression 
models. These approaches employ merely second-moments 
assumptions, allowing to extend the Tweedie regression
models to the class of quasi-Tweedie regression models, which in turn
offer robust and flexible models to deal with continuous data.
Characteristics such as symmetry or asymmetry, heavy tailed and excess 0s 
are easily handled because of the flexibility of the model class. These features
indicate that the Tweedie model is a potential useful tool for 
the modeling of continuous data. The main advantage in practical 
terms, is that we have one model for virtually all kinds of continuous data. 
Thus, model selection is done automatically when fitting the model.

The main advantages of the alternative estimation approaches in relation to
the orthodox maximum likelihood method are their easy implementation
and computational speed. Furthermore, by employing only second-moment
assumptions, we eliminated the non-trivial restriction on the parameter
space of the power parameter, becoming the fitting algorithm simple and
efficient. It also allows us to apply the Tweedie regression 
models for symmetric and heavy tailed data, as the cases of Gaussian and t-Student data, 
where in general the power parameter presents negative and to $0$ values.
Another potential application of Tweedie regression model is for
the analysis of left-skewed data, where we also expect negative values for the 
power parameter.

The theoretical development in Section~\ref{estimation} showed that the
quasi-likelihood approach has much in common with the orthodox maximum 
likelihood method. The quasi-score function employed in the context 
of quasi-likelihood estimation coincides with the score function for 
Tweedie distributions, which also implies that it will coincide for all 
exponential dispersion models. The asymptotic variance of the 
quasi-likelihood estimators for the regression parameters coincide with 
the asymptotic variance of the maximum likelihood estimators, 
in the case of known power and dispersion parameters. 
Hence, the quasi-likelihood approach provides asymptotic efficient estimation
for the regression parameters. Furthermore, the quasi-likelihood approach as 
used in this paper combining the quasi-score and Pearson estimating functions, 
presents the insensitivity property (see Eq. \ref{Sbetalambda}) which is an 
analogue to the orthogonality property in the context of maximum likelihood estimation. 
The insensitive property allows us to apply the two steps Newton scoring algorithm,
using two separate equations to update the regression and dispersion parameters.
A similar procedure can be used in the maximum likelihood framework, since the
vectors $\boldsymbol{\beta}$ and $\boldsymbol{\lambda}$ are orthogonal.
In the context of quasi-likelihood estimation, in this paper, we used the unbiased
Pearson estimating function to estimation of the power and dispersion parameters.
The discussion about efficiency in that case is difficult, since we cannot obtain
a closed form for the Fisher information matrix. The fact that the sensitivity 
and variability matrices associated with the dispersion parameters do not coincide 
indicate that the Pearson estimating functions are not optimum. 
Furthermore, the usage of empirical high-order 
moments for the calculation of the Godambe information matrix must imply some 
efficiency loss. On the other hand, it also becomes the model robust against 
misspecification.

Concerning the pseudo-likelihood approach, it is a well known result that when
$\phi \to 0$ the exponential dispersion models converge to the Gaussian distribution.
Thus, at least for the small variation scenario the Gaussian pseudo-likelihood 
should provide descent estimators for both regression and dispersion parameters.
Furthermore, since the estimators are obtained based on unbiased estimating functions,
we also expect asymptotic unbiased and consistent estimators.
The discussion about efficiency in the context of pseudo-likelihood estimation is difficult,
because of the fact that the regression and dispersion parameters are not orthogonal.
Hence, the asymptotic variance of the regression parameters also depends on high-order
moments. In this paper, we used empirical high-order moments for the calculation of
the asymptotic variance of the pseudo-likelihood estimators. Thus, we expect some
efficiency loss for both regression and dispersion parameters.

The simulation study presented in subsection~\ref{simulation1} showed that in 
general the quasi- and pseudo-likelihood estimators are unbiased and consistent 
for large sample, as suggest the asymptotic results presented in 
Section~\ref{estimation}. In general the coverage rate presented values close
to the nominal level for both methods and simulation scenarios.
The main disadvantage of the quasi- and pseudo-likelihood estimators in
relation to the maximum likelihood is the loss of efficiency on the
estimation of the dispersion parameters, mainly on the high variation
simulation scenario. However, it is important to highlight that the loss
of efficiency on the estimation of the dispersion parameter does not 
affect the efficiency of the regression parameters that in general present
values close to $1$. As expected the maximum likelihood approach did not
work well for small values of the power parameter. The algorithm presented
many convergence problems, mainly when dealing with large sample size.
The simulation study presented in subsection~\ref{simulation2} showed that
at least to some extend the Tweedie regression model can handle heavy tailed
data as generated by the t-Student and slash distributions. 
However, for the cases of high variation data, some loss of efficiency 
on the estimation of the regression parameters is expected.

We illustrated the application of Tweedie regression models through the
analysis of three data sets. The data sets were chosen to cover different types
of continuous data. The first data set illustrates the case of right-skewed and
zero inflation. As expected the three estimation methods estimated the power
parameter in the interval $1$ and $2$, which in turn indicates a compound Poisson
distribution. The second data analyses also deal with right-skewed data, but without
zero inflation, in that case we expected $p \geq 2$. The results of this data analysis
confirmed our expectations. Finally, the third example considered symmetric data. 
In that case, we expected power parameter close to $0$ indicating the Gaussian distribution. 
The two alternative methods confirmed our expectations. 
The maximum likelihood method for this data set converged, since
the sample size is small, but offers a non-optimum fit.
Regarding the estimation in general the quasi-likelihood estimates were 
more similar to the maximum likelihood estimates than the pseudo-likelihood
estimates. In all data analysis the standard errors associated with the
power and dispersion parameters obtained by the alternative methods were
larger than the ones obtained by the maximum likelihood method. It shows the
efficiency loss of these approaches and agrees with the results of our 
simulation study and theoretical development.

Possible topics for further investigation and extensions include
extending the Tweedie regression models to the class of
double Tweedie regression models, where the dispersion
parameter is also described as a function of covariates~\citep{Wu:2015}.
The current version of the fitting algorithms (which is available in on-line supplementary
material) is a preliminary implementation of the Tweedie regression models. 
We plan to develop an \texttt{R} package with a GLM style interface to facilitate and 
propagate the usage of Tweedie regression models. The package should also include
residual analysis and influence measures.

\section*{Acknowledgements}
This paper is dedicated in honor and memory of Professor Bent J{\o}rgensen.
Part of this work was done while the first author was at Laboratory of Mathematics 
of Besan\c{c}on, France. The first author is supported by CAPES (Coordena\c{c}\~ao de
Aperfei\c{c}omento de Pessoal de N\'ivel Superior), Brazil.

\bibliographystyle{dcu} 
\bibliography{BonatKokonendji2016}

\end{document}